\documentclass[10pt,twocolumn]{article}
\usepackage[top=2cm, bottom=2cm, left=1.8cm, right=1.8cm]{geometry}
\usepackage{times}  %
\usepackage[hyphens]{url}
\usepackage{graphicx}  %
\usepackage[keeplastbox]{flushend}
\usepackage[hyphens]{url}  %
\usepackage{graphicx} %
\usepackage{tikzsymbols}

\usepackage{url}                                %
\usepackage{array,multirow,graphicx,adjustbox}  %
\usepackage{booktabs}                           %
\usepackage[utf8]{inputenc}
\usepackage[ruled,algosection,noend,linesnumbered]{algorithm2e}
\usepackage{float}
\usepackage{paralist}
\usepackage{amsmath}
\usepackage{amsfonts}
\usepackage{xcolor}
\usepackage{csquotes} 
\usepackage{tabularx}
\usepackage{makecell}
\usepackage{pbox}
\usepackage[hang,flushmargin]{footmisc}
\usepackage{flushend}
\usepackage{xspace}
\usepackage{multicol, lipsum}
\usepackage[T1]{fontenc}

\usepackage{xcolor}
\usepackage{booktabs}
\usepackage{graphicx}
\usepackage{paralist}
\usepackage[small,bf]{caption}
\usepackage[tight]{subfigure}
\usepackage[hang,flushmargin]{footmisc}

\usepackage[compact]{titlesec}
\titlespacing*{\section}{0pt}{*4}{4pt}
\titlespacing*{\subsection}{0pt}{*3}{3pt}
\usepackage{xspace}

\makeatletter
\def\url@leostyle{%
  \@ifundefined{selectfont}{\def\UrlFont{}}%
  {\def\UrlFont{}}%
}
\makeatother
\usepackage[hyphenbreaks]{breakurl}

\usepackage[bookmarks=true, bookmarksnumbered=true, colorlinks=true, linkcolor=linkcol, citecolor=citecol, urlcolor=urlcol, hypertexnames=true]{hyperref}

\definecolor{darkgreen}{RGB}{0, 100, 0}
\definecolor{linkcol}{rgb}{0.3,0,0}
\definecolor{citecol}{rgb}{0.3,0,0}
\definecolor{urlcol}{rgb}{0.3,0,0}

\makeatletter
\def\url@leostyle{%
  \@ifundefined{selectfont}{\def\UrlFont{\small}}%
  {\def\UrlFont{}}%
}
\makeatother
\newcommand{\descr}[1]{\smallskip\noindent\textbf{#1}}
\newcommand{\dspol}{{{\selectfont /pol/}}\xspace}

\let\OLDthebibliography\thebibliography
\renewcommand\thebibliography[1]{
  \OLDthebibliography{#1}
  \setlength{\parskip}{0pt}
  \setlength{\itemsep}{1pt plus 0.2ex}
}

\newcommand\myninja{\dNinja[1][white][gray]{}}

\newif
\ifcomment
\commenttrue
\ifcomment
\newcommand{\sz}[1]{{\bf \textcolor{brown}{SZ: #1}}}
\newcommand{\edc}[1]{{\bf \textcolor{red}{EDC: #1}}}
\newcommand{\gs}[1]{{\bf \textcolor{green}{GS: #1}}}
\newcommand{\ap}[1]{{\bf \textcolor{blue}{AP: #1}}}
\newcommand{\jbnote}[1]{{\bf \textcolor{magenta}{JB: #1}}}
\else
\newcommand{\sz}[1]{}
\newcommand{\edc}[1]{}
\newcommand{\gs}[1]{}
\newcommand{\ap}[1]{}
\newcommand{\jbnote}[1]{}
\fi

\title{\bf Raiders of the Lost Kek: 3.5 Years of Augmented 4chan Posts\\from the Politically Incorrect Board\thanks{Published at the 14th International AAAI Conference on Web and Social Media (ICWSM 2020). Please cite the ICWSM version.}} 

\author{
Antonis Papasavva\textsuperscript{1,\myninja}, 
Savvas Zannettou\textsuperscript{2,\myninja}, 
Emiliano De Cristofaro\textsuperscript{1,\myninja},\\ 
Gianluca Stringhini\textsuperscript{3,\myninja},
and Jeremy Blackburn\textsuperscript{4,\myninja}\\[0.5ex]
\textsuperscript{1}{University College London}, 
\textsuperscript{2}{Max-Planck-Institut f\"{u}r Informatik},\\
\textsuperscript{3}{Boston University},
\textsuperscript{4}{Binghamton University},
\textsuperscript{\myninja}{iDRAMA Lab}\\
\{antonis.papasavva.19, e.decristofaro\}@ucl.ac.uk, 
szannett@mpi-inf.mpg.de,\\
gian@bu.edu, 
jblackbu@binghamton.edu}
\date{}

\begin{document}
\maketitle

\begin{abstract}
This paper presents a dataset with over 3.3M threads and 134.5M posts from the Politically Incorrect board (\dspol) of the imageboard forum 4chan, posted over a period of almost 3.5 years (June 2016--November 2019).
To the best of our knowledge, this represents the largest publicly available 4chan dataset, providing the community with an archive of posts that have been permanently deleted from 4chan and are otherwise inaccessible.
We augment the data with a set of additional labels, including toxicity scores and the named entities mentioned in each post.
We also present a statistical analysis of the dataset, providing an overview of what researchers interested in using it can expect, as well as a simple content analysis, shedding light on the most prominent discussion topics, the most popular entities mentioned, and the toxicity level of each post.
Overall, we are confident that our work will motivate and assist researchers in studying and understanding 4chan, as well as its role on the greater Web. 
For instance, we hope this dataset may be used for cross-platform studies of social media, as well as being useful for other types of research like natural language processing.
Finally, our dataset can assist qualitative work focusing on in-depth case studies of specific narratives, events, or social theories.

\end{abstract}

\section{Introduction}
Modern society increasingly relies on the Internet for a wide range of tasks, including gathering, sharing, and commenting on content, events, and discussions.
Alas, the Web has also enabled anti-social and toxic behavior to occur at an unprecedented scale.
Malevolent actors routinely exploit social networks to target other users via hate speech and abusive behavior, or spread extremist ideologies~\cite{allison2018social,chetty2018hate,cresci2018fake,waseem2016you}. 

A non-negligible portion of these nefarious activities often originate on ``fringe'' online platforms, e.g., 4chan, 8chan, Gab.
In fact, research has shown how influential 4chan is in spreading disinformation~\cite{kang2016fake,zannettou2017web}, hateful memes~\cite{zannettou2018origins}, and coordinating harassment campaigns on other platforms~\cite{hine2017kek,mariconti2018you,snyder2017fifteen}.
These platforms are also linked to various real-world violent events, including the radicalization of users who committed mass shootings~\cite{christchurch,8chanmanifesto,4chanbellingcat}.

4chan is an imageboard where  
users (aka Original Posters, or OPs) can create a thread by posting an image and a message to a board; others can post in the OP's thread, with a message and/or an image.
Among 4chan's key features are anonymity and ephemerality; users do not need to register to post content, and in fact the overwhelming majority of posts are anonymous.
At most, threads are archived after they become inactive and deleted within 7 days.

Overall, 4chan is widely known for the large amount of content, memes, slang, and Internet culture it has generated over the years~\cite{wp2014understand4chan}.
For example, 4chan popularized the ``lolcat'' meme on the early Web.
More recently, politically charged memes, e.g., ``God Emperor Trump''~\cite{kym2019geotus} have also originated on the platform.

\descr{Data Release.} In this work, we focus on the {\em ``Politically Incorrect'' board} (\dspol),\footnote{\url{http://boards.4chan.org/pol/}} given the interest it has generated in prior research and the influential role it seems to play on the rest of the Web~\cite{pettisambiguity,hine2017kek,zannettou2017web,snyder2017fifteen,zannettou2018origins,tuters2019they}. 
Along with the paper, we release a dataset~\cite{zenodo} including 134.5M posts from over 3.3M \dspol conversation threads, made over a period of approximately 3.5 years (June 2016--November 2019).
Each post in our dataset has the text provided by the poster, along with various post metadata (e.g., post id, time, etc.).

We also {\em augment} the dataset by attaching additional set of labels to each post, including: 1) the named entities mentioned in the post, and 2) the toxicity scores of the post.
For the former, we use the spaCy library~\cite{spacy.io}, and for the latter, Google's Perspective API~\cite{jigsaw2018perspective}. 

We also wish to warn the readers that some of the content in our dataset, as well as in this paper, is highly toxic, racist, and hateful, and can be rather disturbing.

\descr{Relevance.} We are confident that our dataset will be useful to the research community in several ways.
First, \dspol contains a large amount of hate speech and coded language that can be leveraged to establish baseline comparisons, as well as to train classifiers.
Second, due to 4chan's outsized influence on other platforms, our dataset is also useful for understanding flows of information across the greater Web.
Third, our dataset contains numerous events, including highly controversial elections around the world (e.g., the 2016 US Presidential Election, the 2017 French Presidential Election, and the Charlottesville Unite the Right Rally), thus the data can be useful in retrospective analyses of these events.

Fourth, we are releasing this dataset also due to the relatively high bar needed to build a data collection system for 4chan and a desire to increase data accessibility in the community.
Recall that, given 4chan's ephemerality, it is impossible to retrieve old threads.
While there are other, third party archives that maintain deleted 4chan threads, they are either no longer maintained (e.g., \url{chanarchive.org}), are focused around front-end uses (e.g., \url{4plebs}), or are not fully publicly available (e.g., \url{4archive.org}).

\descr{Paper Organization.} The rest of the paper is organized as follows.
First, we provide a high-level explanation on how 4chan works in Section~\ref{sec:whatis4chan}.
Then, we describe our data collection infrastructure (Section~\ref{sec:datacollection}) and 
present the structure of our dataset in Section~\ref{sec:datastructure}.
Next, we provide a statistical analysis of the dataset (Section~\ref{sec:analysis}), followed by a topic detection, entity recognition, and toxicity assessment of the posts in Section~\ref{sec:contentanalysis}.
Finally, after reviewing related work (Section~\ref{sec:relatedwork}), the paper concludes with Section~\ref{sec:conclusion}.

\section{What is 4chan?}\label{sec:whatis4chan}

\begin{figure}[t]
    \centering
    \includegraphics[width=0.99\columnwidth]{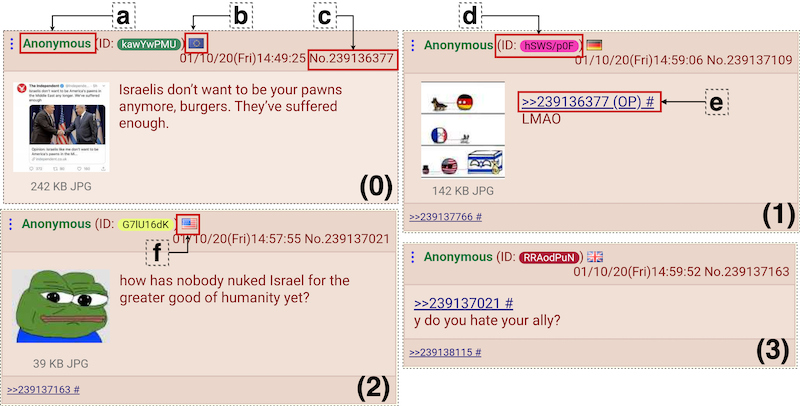}
    \caption{Example of a typical \dspol thread.}
    \label{fig:pol-example-threads}
\end{figure}

\url{4chan.org} is an imageboard launched on October 2003 by Christopher Poole, a then-15-year-old student.
An OP can create a new thread by posting an image and a message to a board. Then, others can post on the OP's thread with a message and/or an image.
Users can also ``reply'' to other posts in a thread by referring to the post ID in their comment.
Figure~\ref{fig:pol-example-threads} shows a typical \dspol thread:
(0) shows the original post, while (1), (2), and (3) are other posts on that thread.

\descr{Boards.}
As of January 2020, 4chan features 70 different boards, which are categorized into 7 high level categories, namely, Japanese Culture, Video Games, Interests, Creative, Other, Misc (NSFW), and Adult (NSFW).
This paper presents a dataset of posts on \dspol, the ``Politically Incorrect'' board, which falls under the Misc category.

\descr{Anonymity.} Users do not need an account to post on 4chan.
When posting, users have the \emph{option} to enter a name along with their post, 
but anonymous posting is the default and by far preferred way of posting on 4chan (see `a' in Figure~\ref{fig:pol-example-threads}).
Note that anonymity in 4chan is meant to be towards other users and not towards the service, as 4chan maintains IP logs and actually makes them available in response to subpoenas~\cite{vice2019arrested}.
Users also have the option to use \emph{Tripcodes}, i.e., adding a password along with a name while posting: the hash of the password will be the unique tripcode of the user, thus making their posts identifiable across threads. 
In addition, some boards, including \dspol, attach a \emph{poster ID} to each post (d in the figure); this is a unique ID linking posts by the same user in the same thread.

\descr{Flags.} Posts on \dspol also include the flag of the country the user posted from, based on IP geo-location.
Obviously, geo-location may be manipulated using VPNs and proxies, however, popular VPNs as well as Tor are blacklisted~\cite{torwiki}.
Note that \dspol is only one of four boards using flags.
Figure~\ref{fig:pol-example-threads} also shows the use of flags on \dspol: the author of post (2) appears to be posting from the US (f).

In addition, users on \dspol can choose \emph{troll flags} when posting, rather than the default geo-localization based country.
As of January 2020 the troll flags options are Anarcho-Capitalist, Anarchist, 
Black Nationalist, Confederate, Communist, Catalonia, Democrat, European, Fascist, Gadsden, Gay, Jihadi, Kekistani, Muslim, National Bolshevik, Nazi, Hippie, Pirate, 
Republican, Templar, Tree Hugger, United Nations, and White Supremacist.
For instance, the OP (post (0)) selected the ``European'' troll flag (b).

\descr{Ephemerality.} Ephemerality is one of the key features of 4chan.
Each board has a limited number of active threads called the \emph{catalog}.
When a user posts to a thread, that thread will be \emph{bumped} to the top of the catalog.

When a new thread is created, the thread at the bottom of the catalog, i.e., the one with the least recent post, is removed.
After the thread is removed from the catalog it is placed into an archive, and then, after 7 days, it is permanently deleted.
That is, popular threads are kept alive by new posts, while less popular threads die off as new threads are created.

However, threads are also limited in the number of times they can be bumped.
When a thread reaches the {\em bump limit} (300 for \dspol), it can no longer be bumped, but does remain active until it falls off the bottom of the catalog.

\descr{Replies.}
Figure~\ref{fig:pol-example-threads} also illustrates the {\em reply} feature of 4chan.
A user can click on the post ID (c) to generate a post including ``\textgreater \textgreater {\em post ID}'' (see, e.g., e in post (1)).

\descr{Moderation.} 4chan has very little moderation, especially on \dspol.
Users can volunteer to be moderators, aka ``janitors.''
Janitors have the ability to delete posts and threads, and also recommend users to be banned.
These recommendations go to 4chan employees who are responsible for reviewing user activity before applying a ban.
Overall, \dspol is considered a containment board, allowing generally distasteful content, even by 4chan standards, to be discussed without disturbing the operations of other boards~\cite{hine2017kek}.

\descr{Slang.} Over the years, 4chan has been the de-facto incubator for a huge number of memes and behaviors that we now consider central to mainstream Internet culture, including lolcats, Rickrolling, and rage comics~\cite{wp2014understand4chan}.
It has also served as a platform for activist movements (e.g., Anonymous) and broad political ideologies like the Alt-Right.
In particular, \dspol discourse is strongly characterized by a rather ``original'' slang, with popular words appearing in our dataset including expressions like 
``Goy'' (a somewhat derogatory term originally used by Jews to denote non-Jews, used on 4chan primarily in reference to anti-Semitic conspiracy theories where Jews act as ``malevolent puppet-masters''~\cite{goyim}), ``Kek'' (which originated as a variant of LOL and became the God of memes, via which they influence reality), ``anon'' (abbreviated for anonymous, describing another 4chan poster), etc.

\section{Data Collection}\label{sec:datacollection}

\begin{table}[t]

\centering
\resizebox{1.0\columnwidth}{!}{
\smallskip\begin{tabular}{l r r r r r}
\toprule
& \textbf{2016} & \textbf{2017} & \textbf{2018} & \textbf{2019} & \textbf{Total} \\
\midrule
\textbf{Threads} & 643,535 & 1,123,341 & 922,103 & 708,932 & 3,397,911\\
\textbf{Posts} & 21,892,815 & 44,573,337 & 39,413,548 & 28,649,533 & 134,529,233 \\
\toprule
\end{tabular}}
\caption{Number of threads and posts in the dataset.}\label{tbl:threadsposts_crawled}
\end{table}

We now discuss our methodology to collect the dataset released along with this paper.

We started crawling \dspol, in June 2016, using 4chan's JSON API.\footnote{\url{https://github.com/4chan/4chan-API}}
(This was done as part of our first academic study of 4chan~\cite{hine2017kek}.) 
Given 4chan's ephemeral nature, we devised the following methodology to ensure we obtained the full/final contents of all threads.
Every 5 minutes, we retrieve \dspol's thread catalog and compare the list of the currently active threads to the ones obtained earlier. 
Once a thread is no longer active, we obtain the full copy of that thread from 4chan's archive.
For each post in a thread, the 4chan API returns, among other things, the post's number, its author, UNIX timestamp, and content of the post.
We explain in detail our dataset and what it contains in the 
next section.
Note that while we do not provide posted images, posts do include image metadata, e.g., filename, dimensions (width and height), file size, and an MD5 hash of the image.

Table~\ref{tbl:threadsposts_crawled} provides an overview of our dataset. 
Note that for, about $6\%$ of the threads, the crawler gets a 404 error: from a manual inspection, it seems that this is due to ``janitors'' (i.e., volunteer moderators) removing threads for violating rules.

The data released with this paper, as well as the analysis presented in later sections, spans from June 29, 2016 to November 1, 2019.
Alas, our dataset has some (minor) gaps due to failure of our data collection infrastructure; specifically, we are missing 10, 4, and 8 days worth of posts during 2016 (October 15 and December 16--24), 2017 (January 10--12 and May 13), and 2019 (April 13 and July 21--27).

\descr{Ethical considerations.} 4chan posts are typically anonymous, however, analysis of the activity generated by links on 4chan to other services could be potentially used to de-anonymize users.
Overall, we followed standard ethical guidelines~\cite{rivers2014ethical} and made no attempt to de-anonymize users. 
Also note that the collection and release of this data does not violate 4chan's API Terms of Service.

\section{Data Structure}\label{sec:datastructure}
\begin{figure}[t]
    \centering
    \includegraphics[width=0.95\columnwidth]{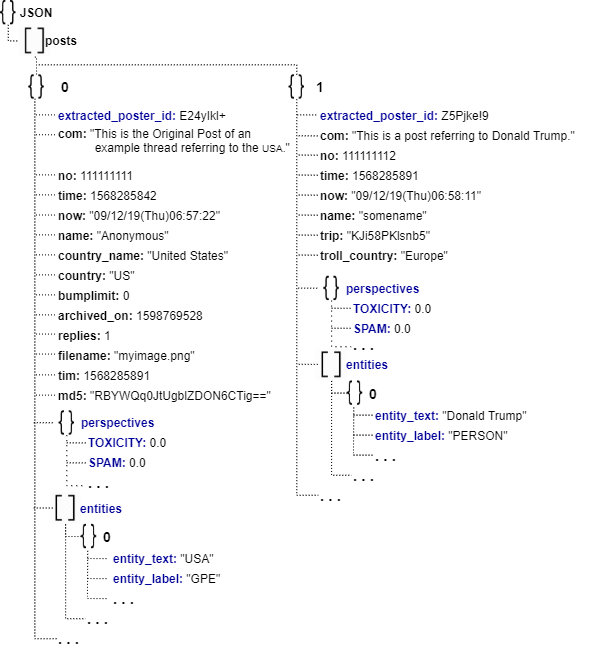}
    \caption{Schematic representation of the JSON structure of the threads in our dataset. (Some keys are omitted to ease presentation.)}
    \label{fig:json_structure}   
\end{figure}

In this section, we present the structure of our dataset, available from~\cite{zenodo}.

The dataset is released as a single newline-delimited JSON\footnote{\url{http://ndjson.org/}} file (\texttt{.ndjson}), with each line consisting of a full thread.
More specifically, each line is a JSON object which contains a list of posts from a single thread.
Each post is a JSON object containing all the key/values returned by the 4chan API, along with three additional ones ({\em entities}, {\em perspectives}, and {\em extracted\_poster\_id}); see below.
Note that the poster ID (d in Figure~\ref{fig:pol-example-threads}) is not always available from the 4chan API.
As of this writing, the API does not return poster IDs for archived threads, but at certain points of our collection period, it did.
To ensure that our dataset includes the poster ID our data collection infrastructure parses the HTML catalog of the 4chan threads to capture it and store it with the key {\em extracted\_poster\_id}:  $95\%$ of the posts have an extracted\_poster\_id.

In Figure~\ref{fig:json_structure}, we report the JSON structure of a thread with two posts: the original post and the second post, with index 0 and 1, respectively.
Due to space limitations, we only list some of the keys, i.e., the most relevant to the analysis presented in the rest of the paper. 
The complete list of keys, along with the type of values they hold and any related documentation, is available at~\cite{zenodo}.

\descr{Keys/Values from the API.} Each post includes the following key/values:
\begin{compactitem}[--]
\item \emph{extracted\_poster\_id}: the poster ID.
\item \emph{com}: the post text in HTML escaped format.
\item \emph{no}: the numeric (unique) post ID.
\item \emph{time}: UNIX timestamp of the post.
\item \emph{now}: human-readable format of the UNIX timestamp.
\item \emph{name}: the name of the poster (default to ``Anonymous'').
\item \emph{trip}: a unique ID to the poster, a hash computed based on the password provided by the user, if any.
\item \emph{country\_name}: full name of the country the user posts from.
\item \emph{country}: country code in Alpha ISO-2 format.
\item \emph{troll\_country}: the troll flag selected by the poster, if any.
\item \emph{bumplimit} (only in the original post): flag indicating whether a thread reached the board's bump limit.
\item \emph{archived\_on} (only in the original post): UNIX timestamp of the time the thread is archived.
\item \emph{replies} (only in the original post): the number of posts the thread has, without counting the original post. 
\end{compactitem}

\smallskip\noindent As mentioned, we do not crawl images, however, the 4chan API returns some image metadata, e.g.;
\begin{compactitem}[--]
\item \emph{filename}: image name as stored on poster's device.
\item \emph{tim}: the time the image is uploaded as a UNIX timestamp.
\item \emph{md5}: the MD5 hash of the image. Note that the image can be found, using the MD5 hash, in unofficial 4chan archives like 4plebs.\footnote{\url{https://4plebs.org/}}
\end{compactitem}

\descr{Named Entities.}
For each JSON object, we complement the data with the list of the named entities we detect for each post, using the spaCy (v2.2+) Python library~\cite{spacy.io}.
For each entity, we include a dictionary with four different characteristics of the named entity, namely: 
\begin{compactitem}[--]
\item \emph{entity\_text}: the name of the detected entity. 
\item \emph{entity\_label}: the type of the named entity. %
\item \emph{entity\_start}: character index in \emph{com} in which the named entity starts.
\item \emph{entity\_end}: character index in \emph{com} in which the named entity ends.
\end{compactitem}

\descr{Perspective Scores.} We also add scores returned by the Google's Perspective API~\cite{jigsaw2018perspective}, and more specifically seven scores in the $[0,1]$ interval:
\begin{compactitem}[--]
\item {\sc toxicity} (v6)
\item {\sc severe\_toxicity} (v2)
\item {\sc inflammatory} (v2)
\item {\sc profanity} (v2)
\item {\sc insult} (v2)
\item {\sc obscene} (v2)
\item {\sc spam} (v1)
\end{compactitem}

The process of augmenting every post in our dataset with the named entities and the perspective scores took place between January 2--9, 2020.

\descr{FAIR Principles.} The data released along with this paper aligns with the FAIR guiding principles for scientific data.\footnote{\url{https://www.go-fair.org/fair-principles/}} 
First, we make our data \emph{Findable} by assigning a unique and persistent digital object identifier (DOI): 10.5281/zenodo.3606810.\footnote{\url{https://doi.org/10.5281/zenodo.3606810}}
Second, our dataset is \emph{Accessible} as it can be downloaded, for free, and is in the standard JSON format.
JSON is widely used for storing data and has an extensive and detailed documentation for all of the computer programming languages that support it, thus enabling our data to be \emph{Interoperable}.
Finally, our dataset comes with rich metadata that are extensively documented and described in this paper, in~\cite{zenodo}, and in the 4chan API documentation as well.
The data is released in full and hence is \emph{Reusable}.

\section{General Characterization}\label{sec:analysis}

\begin{figure}[t!]
\centering
\subfigure[Threads]{\includegraphics[width=1\columnwidth]{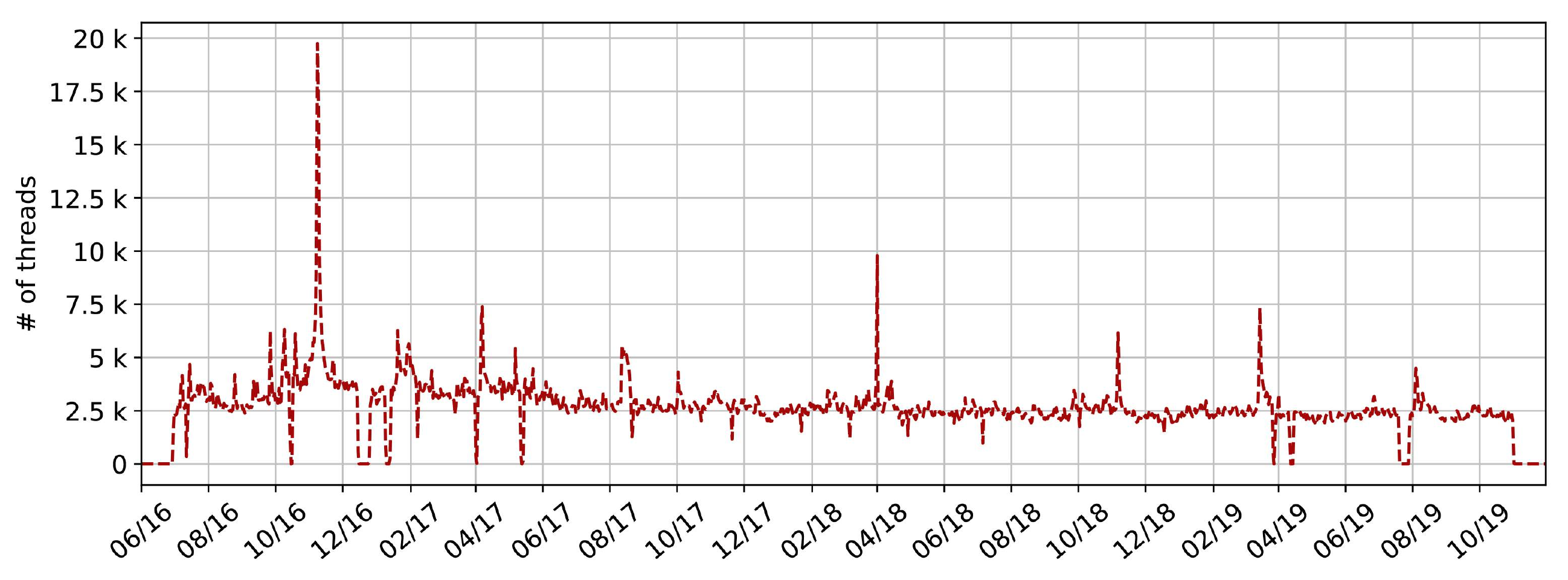}\label{fig:oneline_threads_per_day}}
\subfigure[Posts]{\includegraphics[width=1\columnwidth]{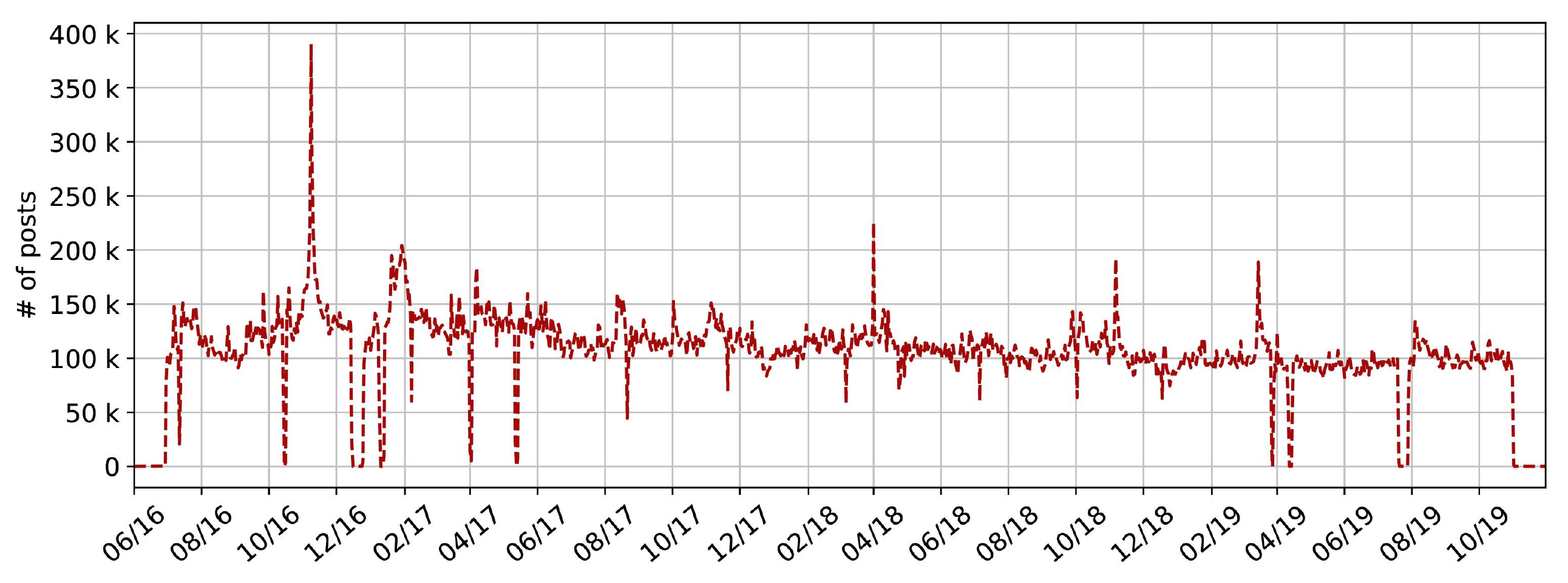}\label{fig:oneline_posts_per_day}}
\caption{Number of threads and posts shared per day.}
\label{fig:oneline_threads_posts_per_day}
\end{figure}

In this section, we provide a general characterization of the dataset that we release.
Our dataset spans 3.5 years, and this prompts the need to shed light on the temporal evolution of \dspol.
Moreover, we analyze the use of tripcodes, images, and flags within the board, aiming to showcase some of the peculiar features that characterize 4chan.

\descr{Posting Activity.} We start by looking at how \dspol's posts are shared over time.
Figure~\ref{fig:oneline_threads_per_day} and Figure~\ref{fig:oneline_posts_per_day} show the number of threads and posts created per day, respectively.
On average, throughout our dataset, over 2.8K threads and 112.3K posts are posted every day on the board.
We observe a peak in posting activity on November 5-13, 2016 (around the US Presidential Election) with 390K posts just on November 8 (Election Day), followed by another peak that lasts from January 20 (Donald Trump's inauguration: 195K posts) until February 3, 2017. Notably, the highest number of posts between these two weeks is observed on January 29 with 204K posts when Donald Trump issued a 90-day travel ban for certain nationals~\cite{dailymail2017bans}.
Additional peaks can be observed close to other world events: 
(1) on April 7, 2017 (184K posts) when Donald Trump ordered missile strikes in Syria~\cite{missile2017nytimes};
(2) on April 1, 2018 (225K posts), possibly due to Donald Trump criticizing California's Governor Jerry Brown's decision to grant 56 pardons~\cite{theweek2019things}; (3) on November 6, 2018 (192K posts), when the US Midterm Election took place;
and (4) March 15, 2019 (189K posts), when 
51 people died in a terrorist attack in a New Zealand mosque~\cite{terrorist2019guardian}. 

Overall, posting activity on \dspol is strongly related to important events worldwide and is known to spread conspiracy theories after catastrophic events take place. 
Notably, numerous mainstream news outlets point to 4chan as the conspiracy theory originator; for instance, about the phrase ``cheese pizza'' referring to a pedophilic code in Hilary Clinton's leaked emails~\cite{newstatesman2019pizzagate}, the ``deep state'' organization against Donald Trump's administration~\cite{guardian2018qanon}, or about the Notre Dame fire~\cite{newstatesman2019notredame}.
Therefore, we are confident our dataset will be useful for further research analyzing conversations on 4chan, as well as activity within and spilling off the platform in response to important events and breaking news.

\begin{figure}[t!]
\centering
\hspace{-0.25cm}
\subfigure[Threads -- Hour of the week]{\includegraphics[width=0.24\textwidth]{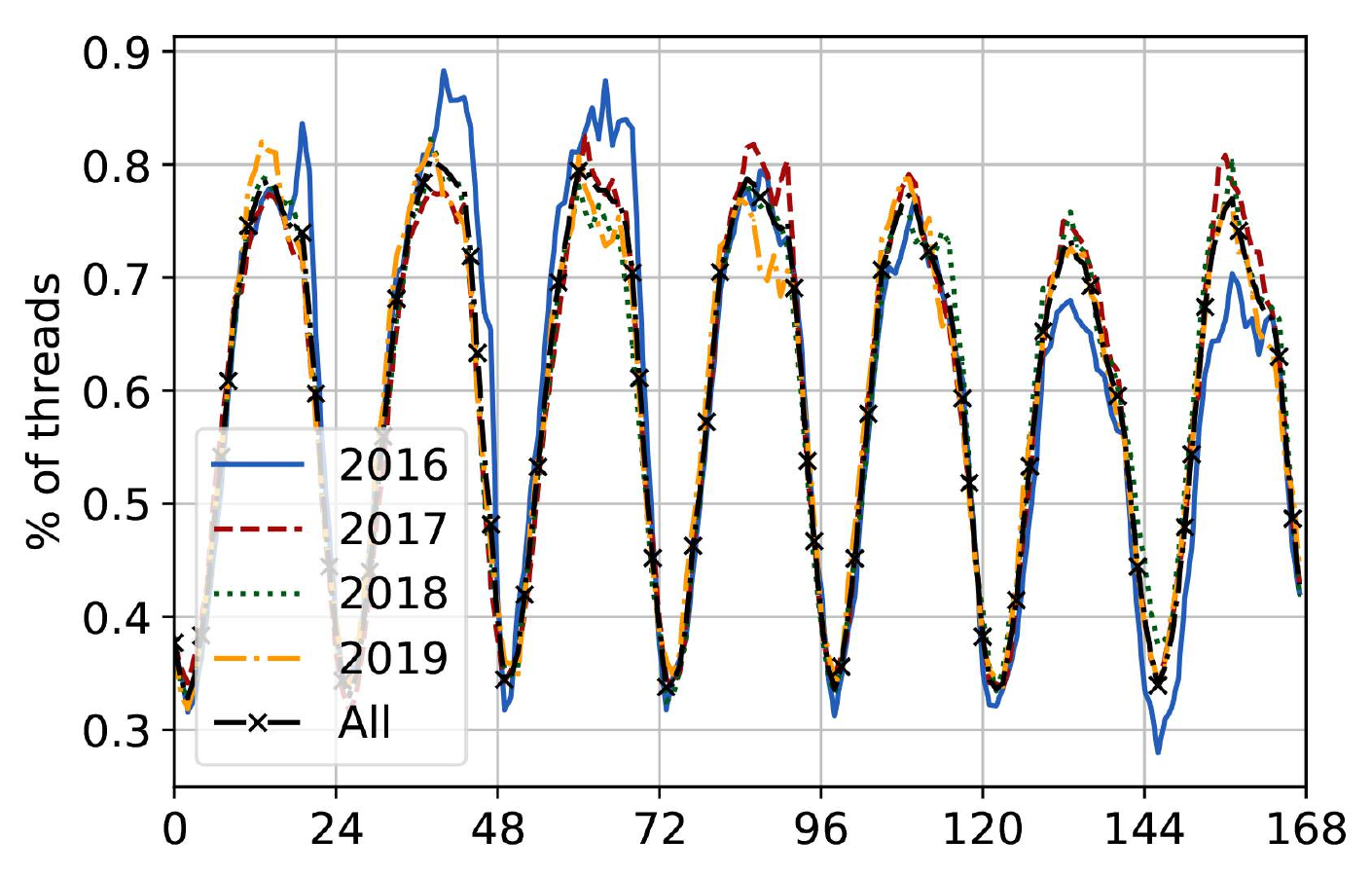}\label{fig:threads_per_hour_week_per_year}}\hspace{-0.25cm}
\subfigure[Posts  -- Hour of the week]{\includegraphics[width=0.24\textwidth]{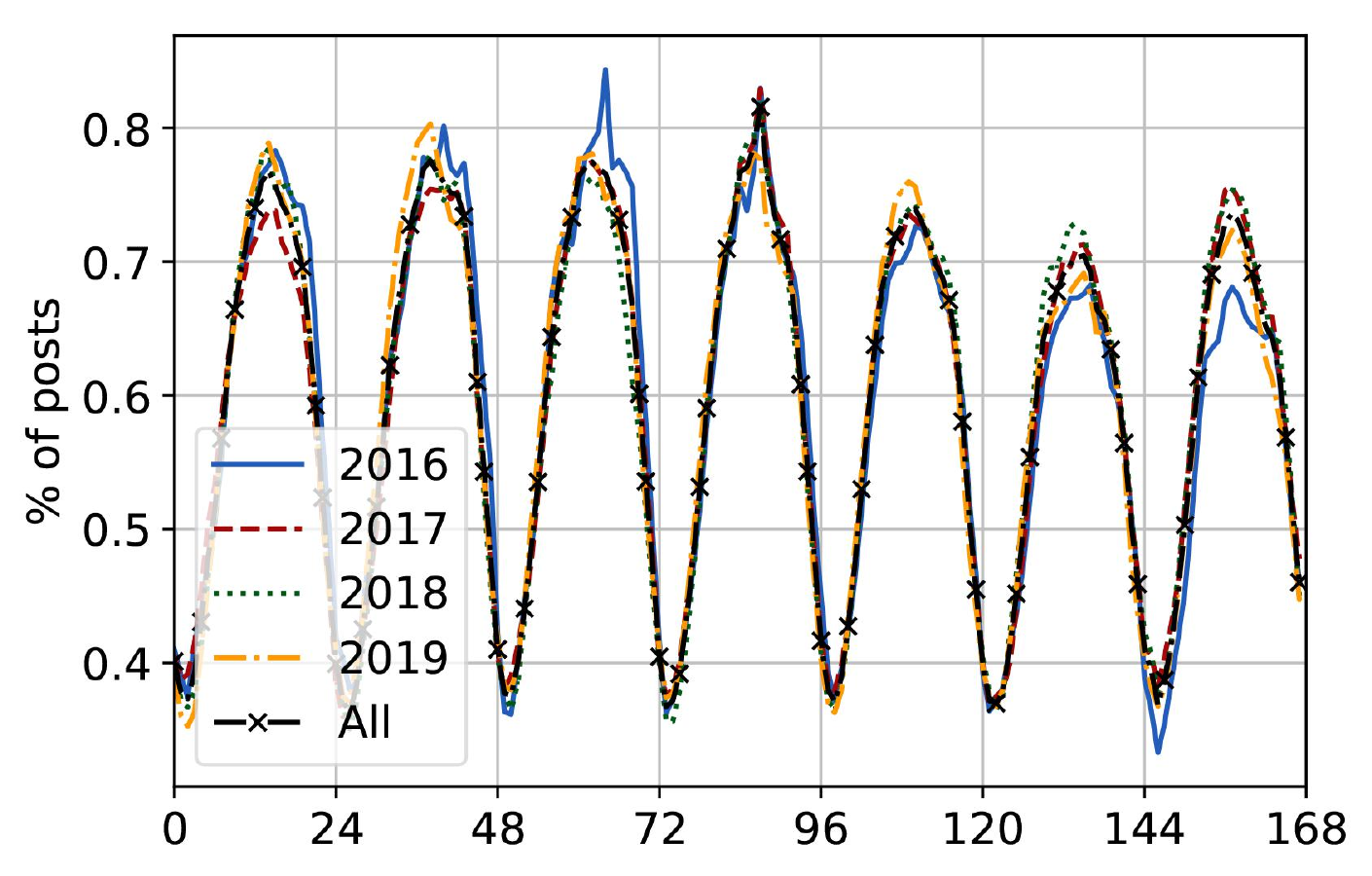}\label{fig:posts_per_hour_week_per_year}}\\
\hspace{-0.25cm}
\subfigure[Threads -- Hour of the day]{\includegraphics[width=0.24\textwidth]{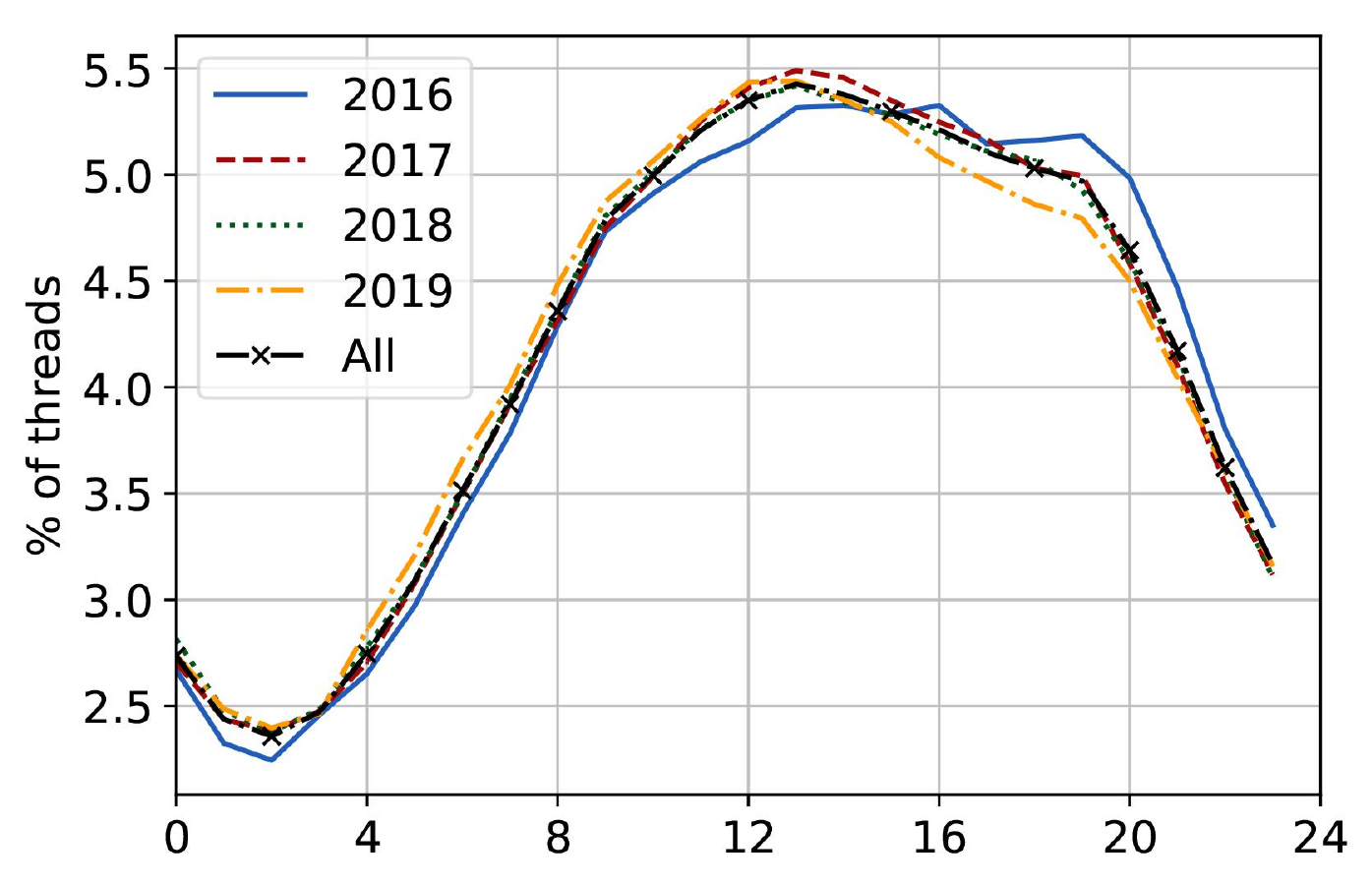}\label{fig:threads_per_hour_day_per_year}}
\hspace{-0.25cm}
\subfigure[Posts -- Hour of the day]{\includegraphics[width=0.24\textwidth]{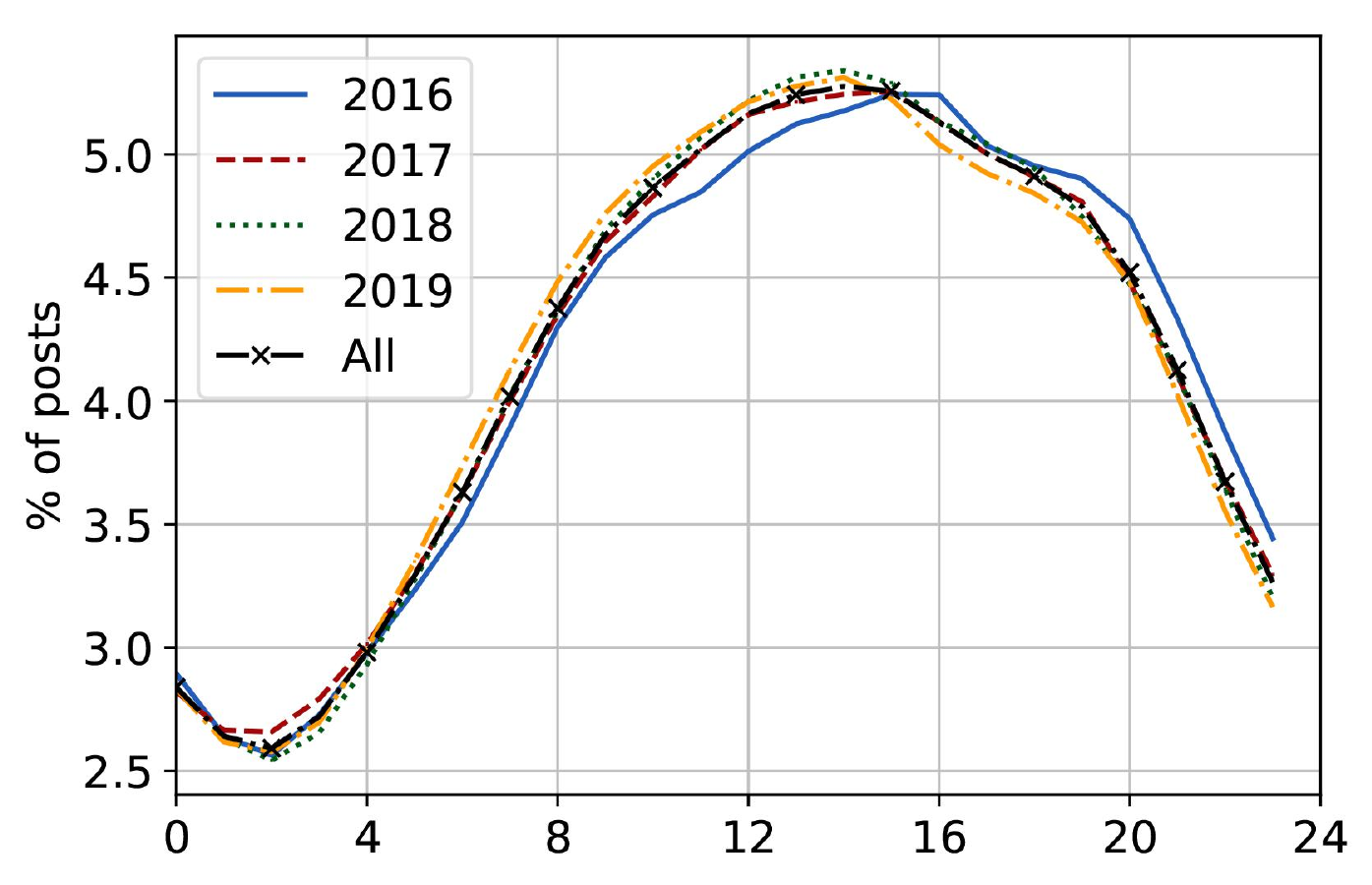}\label{fig:posts_per_hour_day_per_year}}
\caption{Temporal characteristics of threads/posts per hour of week and day. (UTC time zone, week starts on Monday.)}
\label{fig:hour_day_week}
\end{figure}

\descr{Temporal Patterns.} We also look for temporal patterns throughout the day/week.
In Figure~\ref{fig:hour_day_week}, we report the percentage of threads and posts, as per hour of day as well as hour of week.
We do so comparing across the years, finding a very similar behavior throughout.
Overall, we observe that the activity seems to peak during what appear to be the hours of the day in Western countries and more or less weekdays. 

\begin{figure}[t!]
\centering
\hspace{-0.25cm}
\subfigure[\#threads created per flag]{\includegraphics[width=0.24\textwidth]{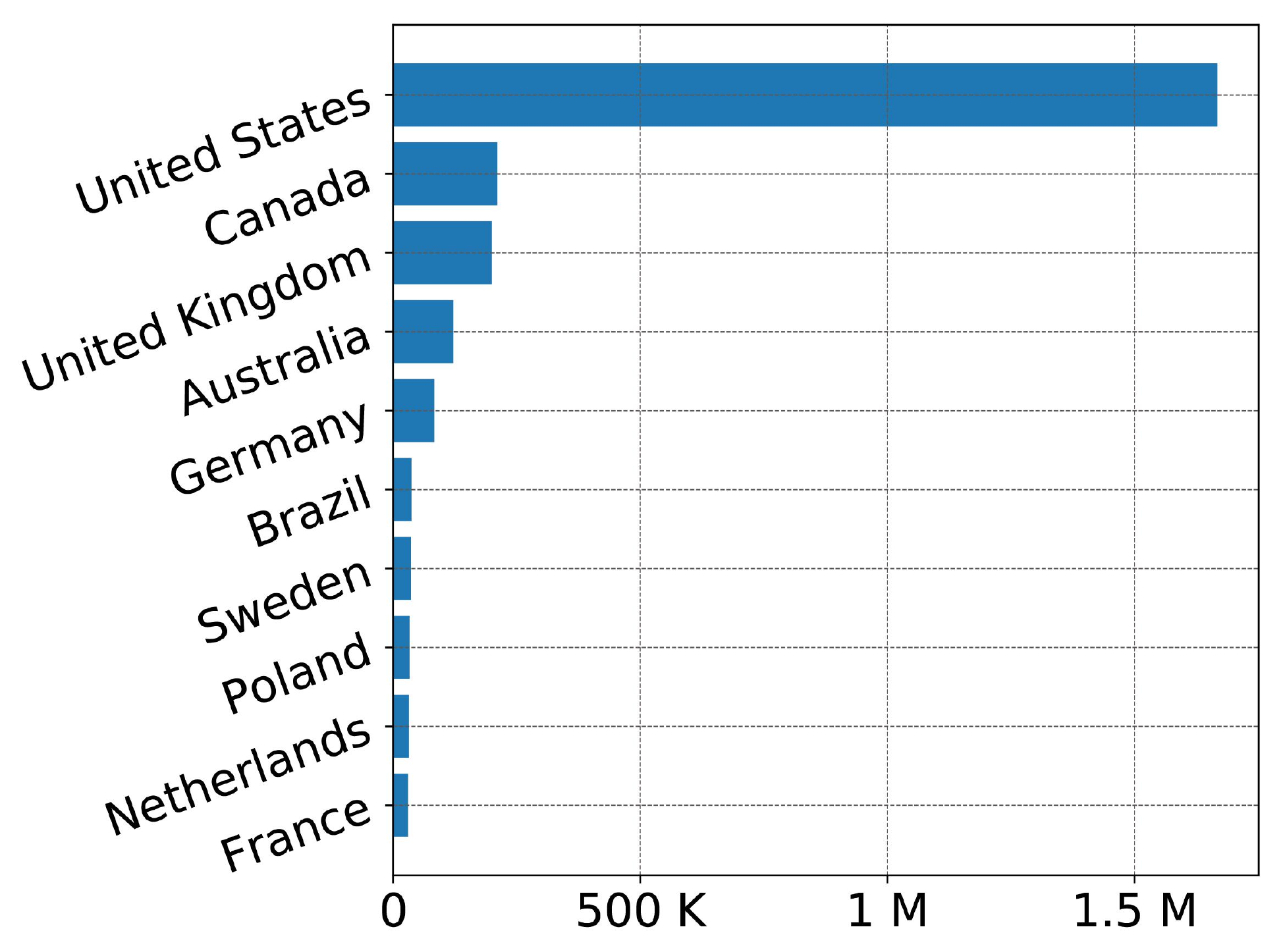}\label{fig:threads_per_flag}}
\hspace{-0.25cm}
\subfigure[\#posts per flag]{\includegraphics[width=0.24\textwidth]{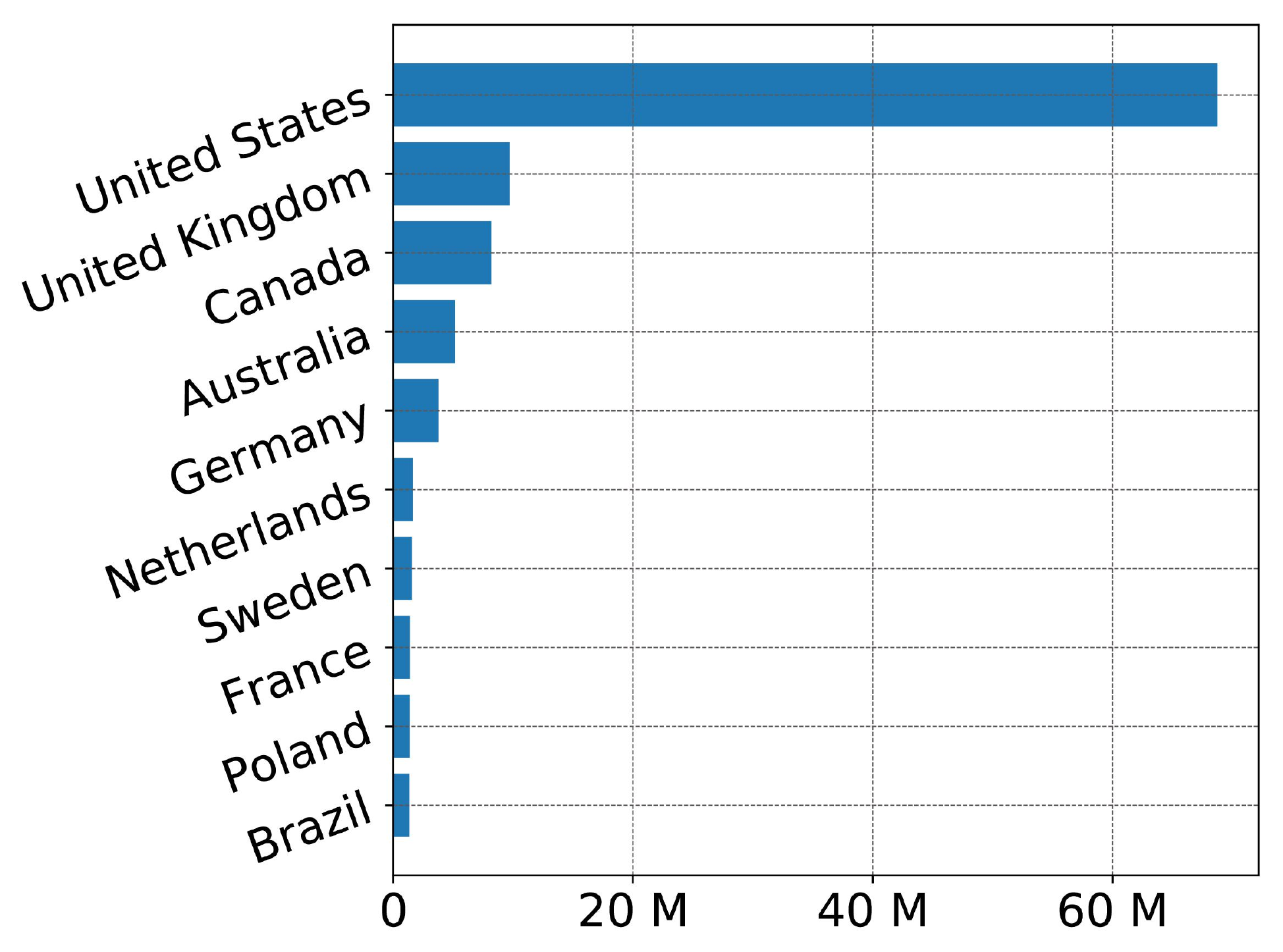}\label{fig:posts_per_flag}}\\
\hspace{-0.25cm}
\subfigure[\#threads created per troll flag]{\includegraphics[width=0.24\textwidth]{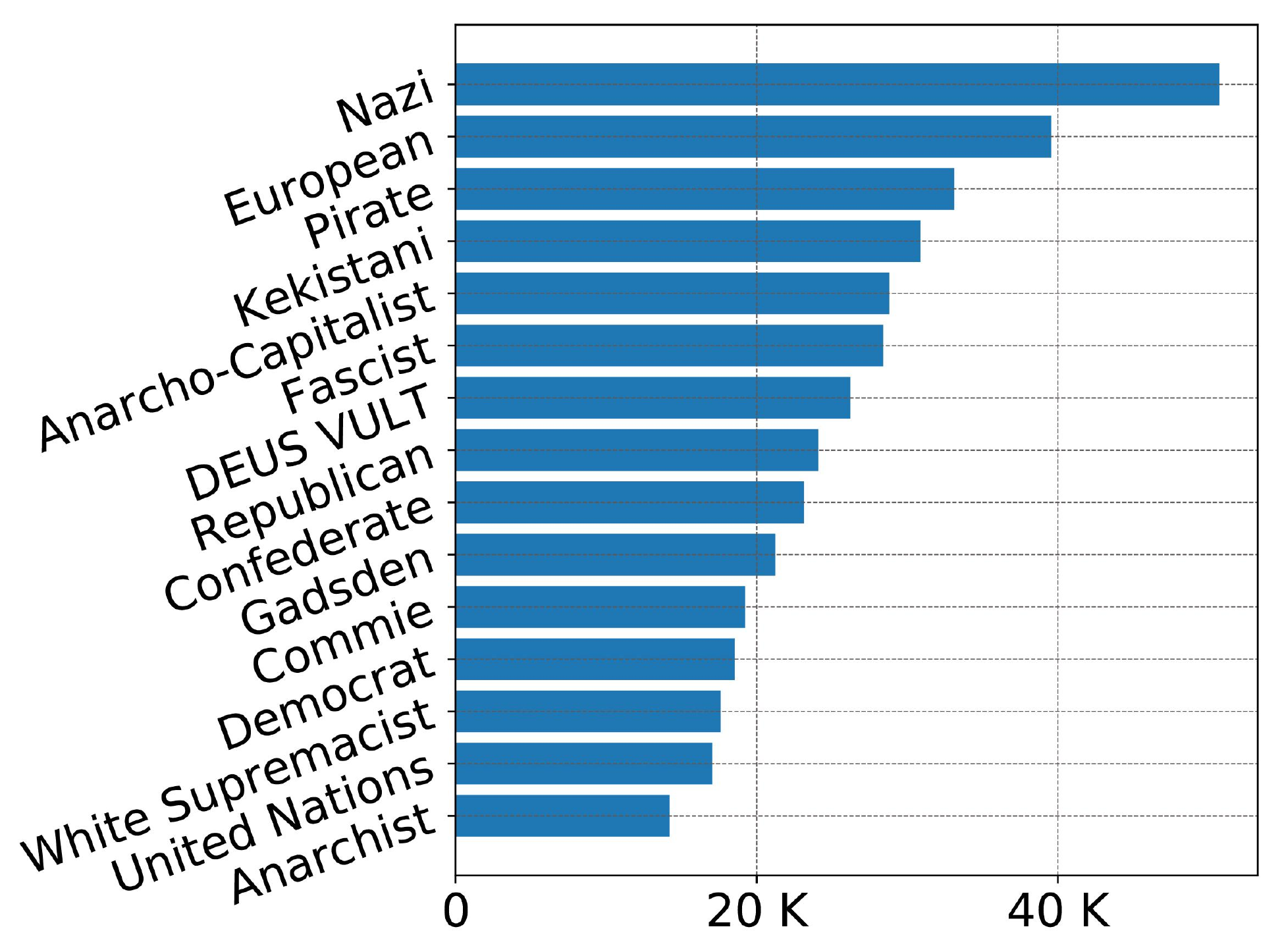}\label{fig:threads_per_trollflag}}
\hspace{-0.25cm}
\subfigure[\#posts per troll flag]{\includegraphics[width=0.24\textwidth]{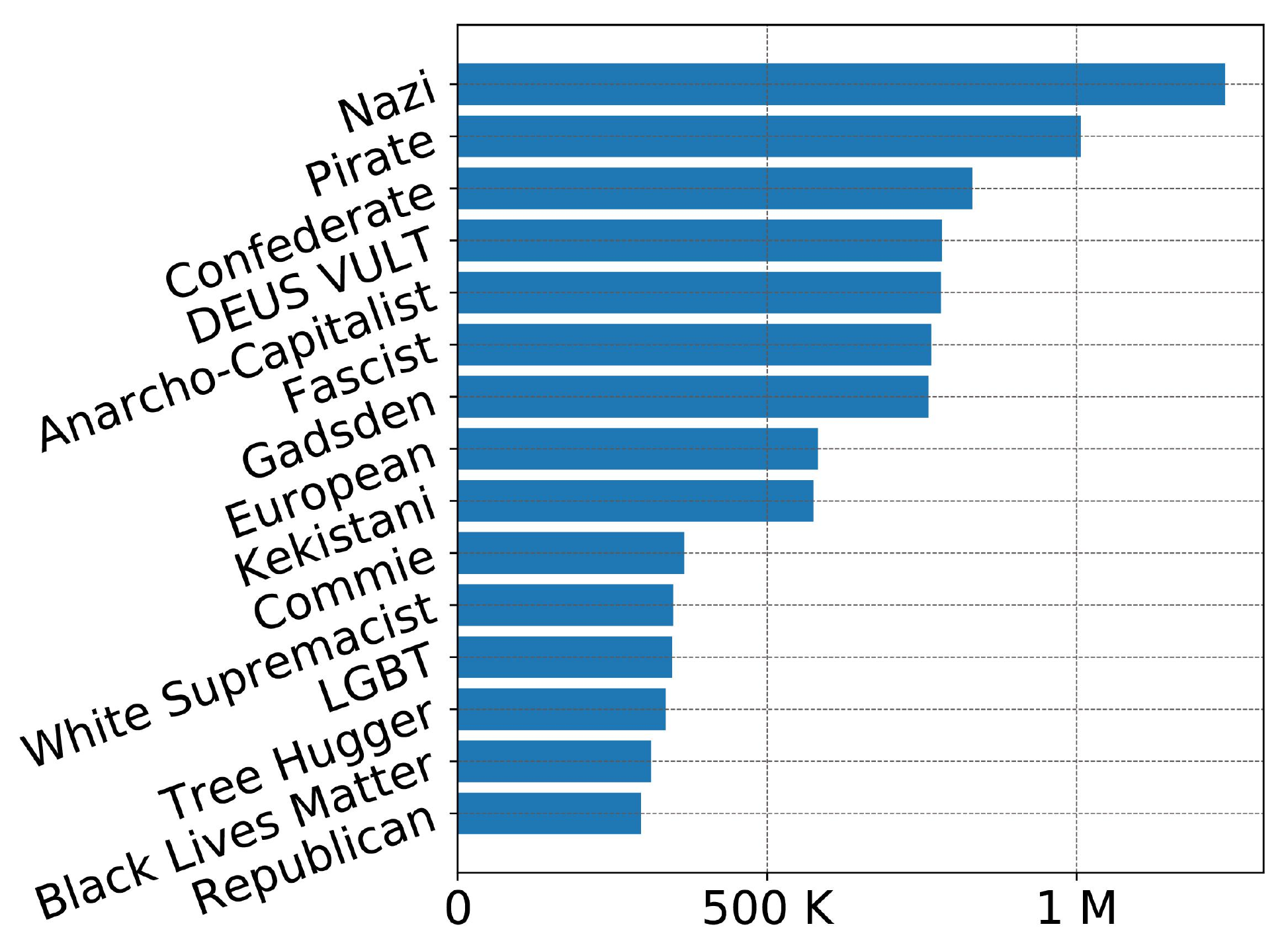}\label{fig:posts_per_trollflag}}
\caption{Number of threads created and posts per flag and troll flag.}
\label{fig:threads_posts_flags}
\end{figure}

\descr{Flags.} We then look at the countries where posts originate, using the flags displayed on \dspol. 
Recall that these are based on IP geo-localization so at best they provide a {\em signal} for general trends and should not be taken at face value.
In Figure~\ref{fig:threads_posts_flags}, we report the top 10 countries, along with the number of threads (Figure~\ref{fig:threads_per_flag}) and overall posts (Figure~\ref{fig:posts_per_flag}) they created.
The most active countries are the US (1.6M threads and 68M posts), followed by the UK (200K threads and 9.7M posts), Canada (210K threads and 8.1M posts), Australia (121K threads and 5.1M posts), and Germany (83.3K threads and 3.7M posts).
We also report the top 15 ``troll flags'' with ``Nazi'' being the most popular with over 50K threads (Figure~\ref{fig:threads_per_trollflag}) and 1.2M posts (Figure~\ref{fig:posts_per_trollflag}). 

\begin{figure}[t!]
\centering
\subfigure[threads]{\includegraphics[width=0.99\columnwidth]{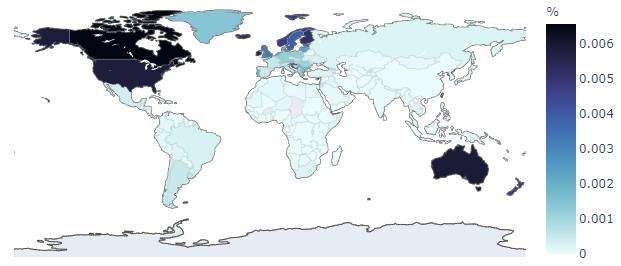}\label{fig:threads_choropleth}}
\subfigure[posts]{\includegraphics[width=\columnwidth]{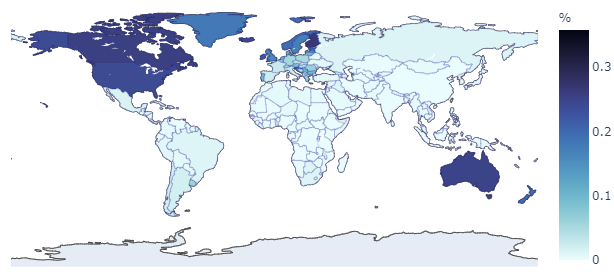}\label{fig:posts_choropleth}}
\caption{Choropleth of the number of threads created/posts per country, normalized by Internet-using population. }
\label{fig:thread_post_choropleth}
\end{figure}

Figure~\ref{fig:threads_choropleth} and~\ref{fig:posts_choropleth}, depict the choropleths of the number of threads and posts created per country worldwide, respectively, this time {\em normalized} using each country's estimated Internet-using population.\footnote{\url{https://www.internetlivestats.com/internet-users-by-country/}} 
While the US dominates in terms of sheer volume of threads created (Figure~\ref{fig:threads_per_flag}), when taking into account the number of Internet users, the top 5 countries actually are Canada (0.0066), Australia (0.0059), US (0.0058), Ireland (0.0058), and Croatia (0.0054).
As for posts, the top 5 countries are Monaco (0.35), Finland (0.26), Canada (0.25), Australia (0.25), and Iceland (0.24).
Overall,
besides Croatia, Monaco, and Finland, we find a number of North and East European countries being relatively active.

\begin{figure}[t!]
\centering
\subfigure[]{\includegraphics[width=0.49\columnwidth]{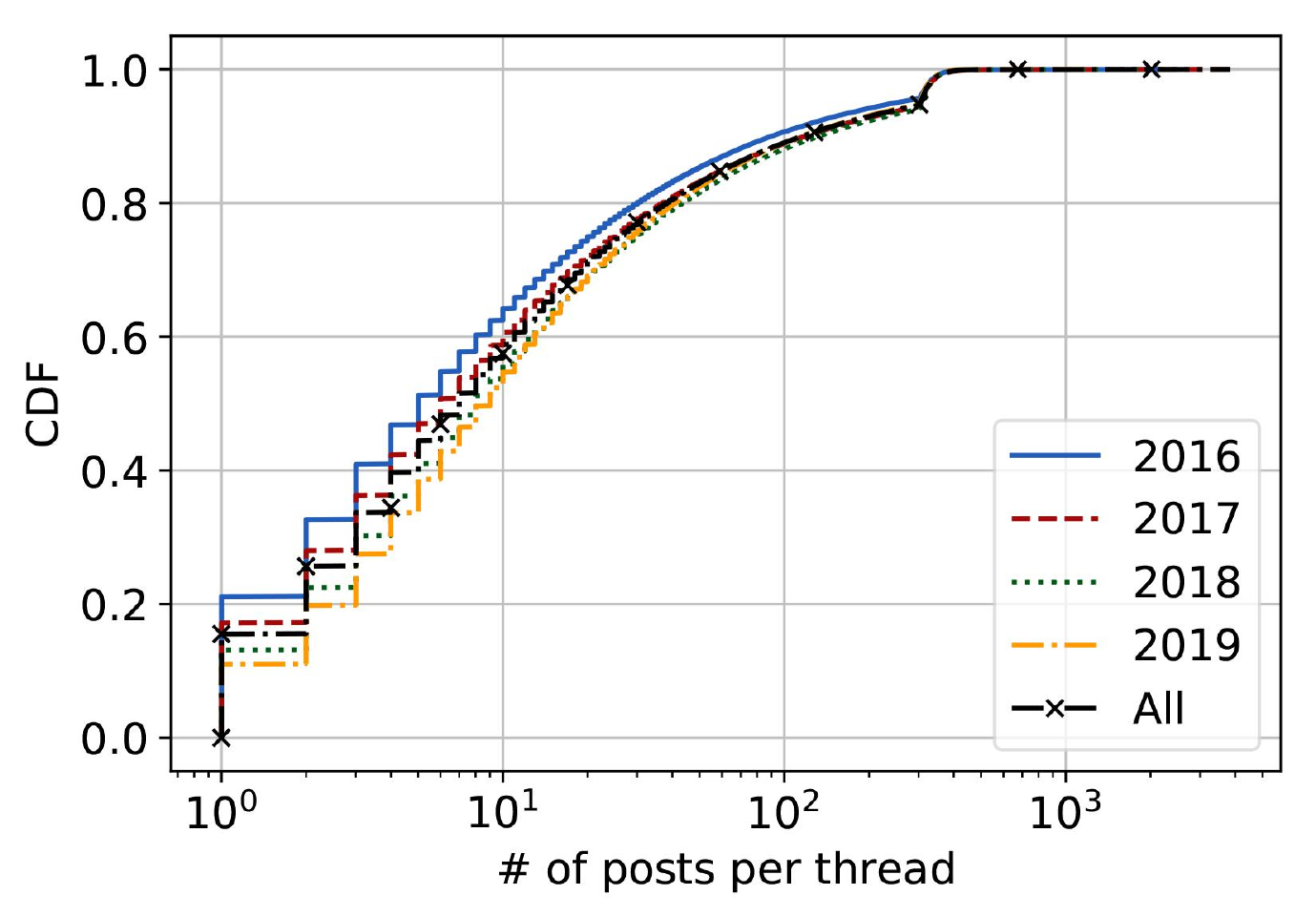}\label{fig:cdf_posts_per_thread_per_year}}
\subfigure[]{\includegraphics[width=0.49\columnwidth]{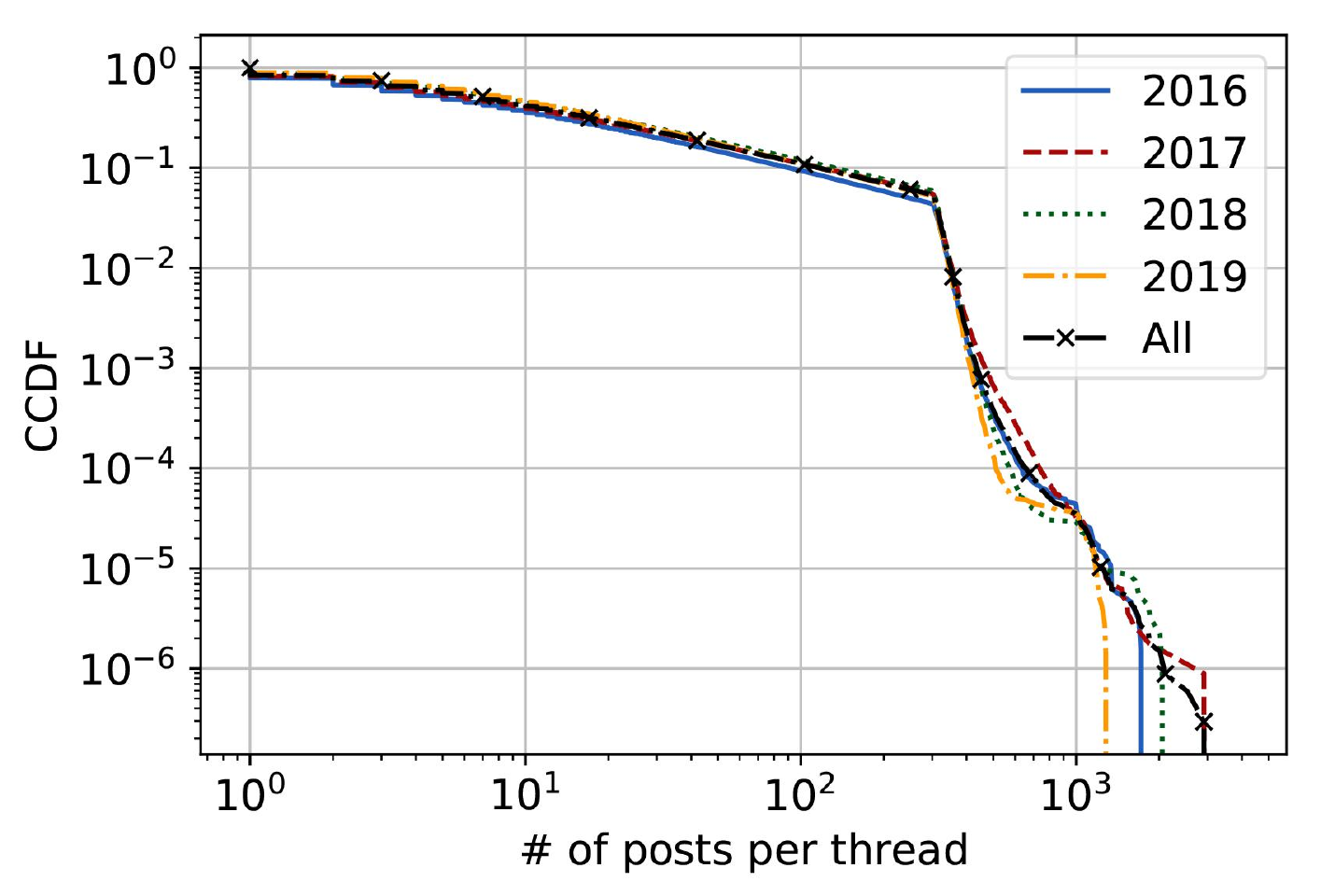}\label{fig:ccdf_posts_per_thread_per_year}}
\caption{CDF and CCDF of the number of posts per thread.}
\label{fig:cdf_ccdf_posts_per_thread}
\end{figure}

\descr{Thread Engagement.} Next, we look at how many posts threads tend to get. 
On average, there are 39.6 posts per thread throughout our dataset, with this number increasing over the years, and specifically 34, 39.7, 42.7, and 40.4 for 2016, 2017, 2018, 2019, respectively.
To capture the distribution of posts per threads we plot the Cumulative Distribution Function (CDF) and the Complementary Cumulative Distribution Function (CCDF) for each year in Figure~\ref{fig:cdf_ccdf_posts_per_thread}. 
The figure highlights that, overall, more \dspol threads tend to get more posts over time.
Specifically, $37\%$, $41\%$, $43\%$, and $44\%$ of the threads in 2016, 2017, 2018, and 2019, respectively, have over 100 posts.

We also test for statistically significant differences between the distributions, using a two-sample Kolmogorov-Smirnov (KS) test, finding them on each pair ($p<0.01$). 
Thus, this suggests that the change over the year is indeed significant.

\descr{Tripcodes.} Next, we study the use of tripcodes by \dspol users to see whether this is negligible or relatively widespread.
Recall that tripcodes are the only way a user can ``sign'' their posts on 4chan, letting others recognize posts made by the same user across different threads.
For instance, the QAnon far-right conspiracy theory (built around alleged efforts by the ``deep state'' against US President Donald Trump) started with a post on 4chan in October 2017 by someone using the name Q~\cite{guardian2018qanon}; Q has reportedly used tripcodes on 4chan and 8chan to ``authenticate'' themselves.

\begin{figure}[t!]
\centering
\subfigure[]{\includegraphics[width=0.49\columnwidth]{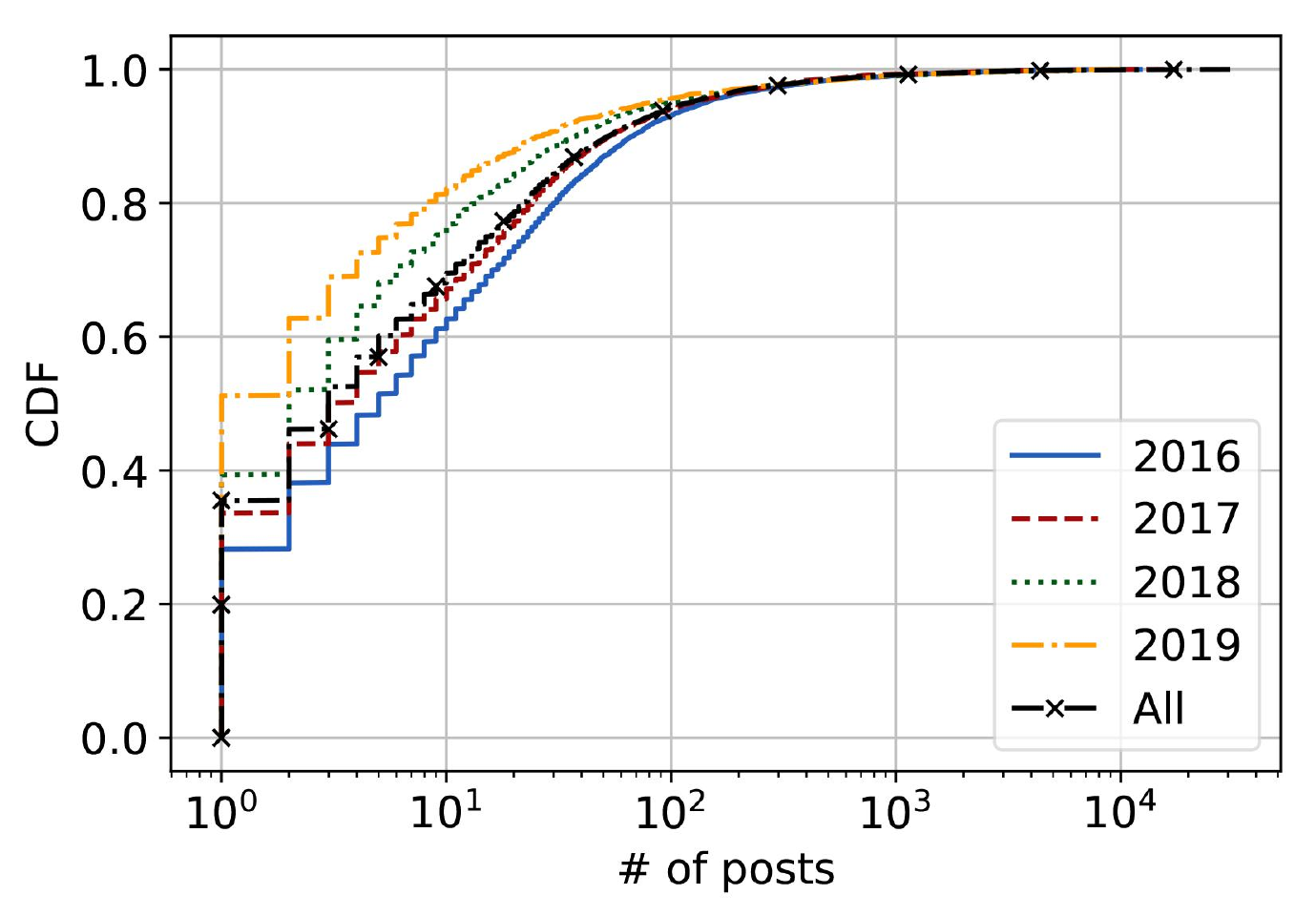}\label{fig:cdf_post_per_year_tripcode.pdf}}
\subfigure[]{\includegraphics[width=0.49\columnwidth]{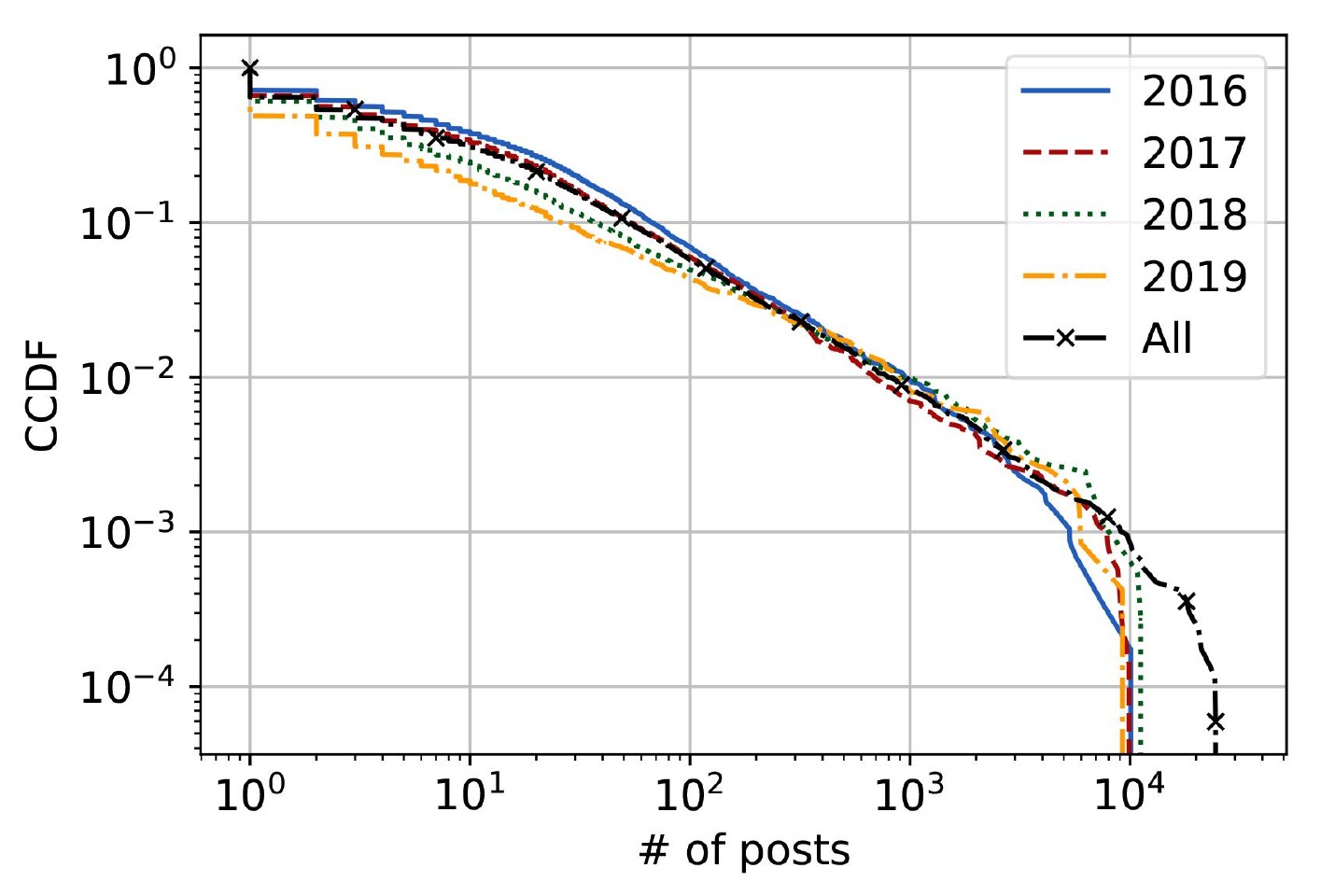}\label{fig:ccdf_post_per_year_tripcode.pdf}}
\caption{CDF and CCDF of the number of posts with unique tripcode.}
\label{fig:cdf_ccdf_posts_per_tripcode}
\end{figure}

In Figure~\ref{fig:cdf_ccdf_posts_per_tripcode}, we plot the CDF and the CCDF of the number of posts with unique tripcode.
Overall, we find that the use of tripcodes goes down over the years.
\begin{compactitem}[--]
\item 2016: 311K posts ($0.23\%$) with unique tripcode from 5.7K different posters;
\item 2017: 365.6K posts ($0.27\%$) from 7.1K posters;
\item 2018: 206K posts ($0.15\%$) from 3.6K posters;
\item 2019: 117K posts ($0.09\%$) from 2.3K posters.
\end{compactitem}

\begin{figure}[t!]
\centering
\subfigure[]{\includegraphics[width=0.49\columnwidth]{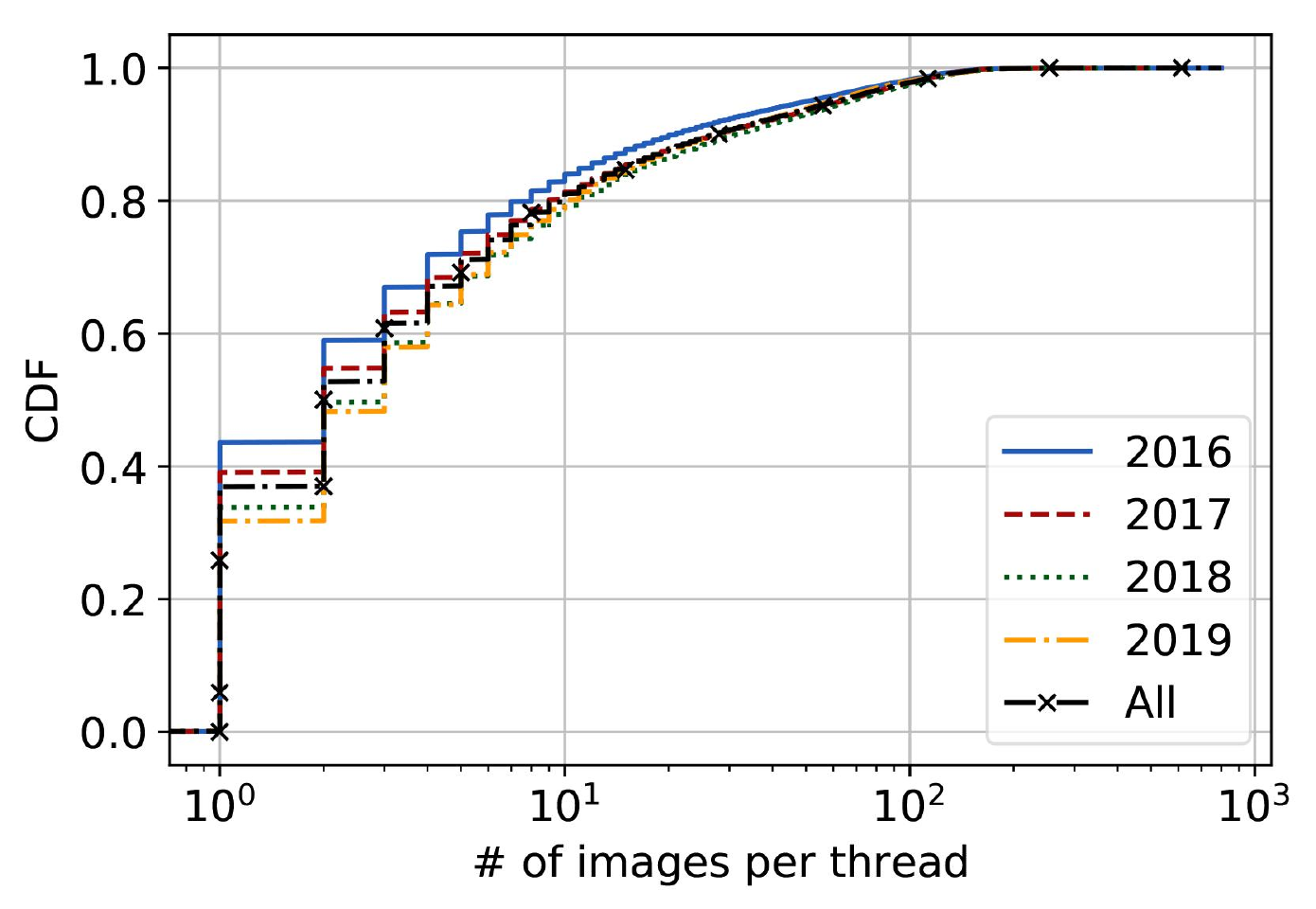}\label{fig:cdf_pictures_per_thread_per_year}}
\subfigure[]{\includegraphics[width=0.49\columnwidth]{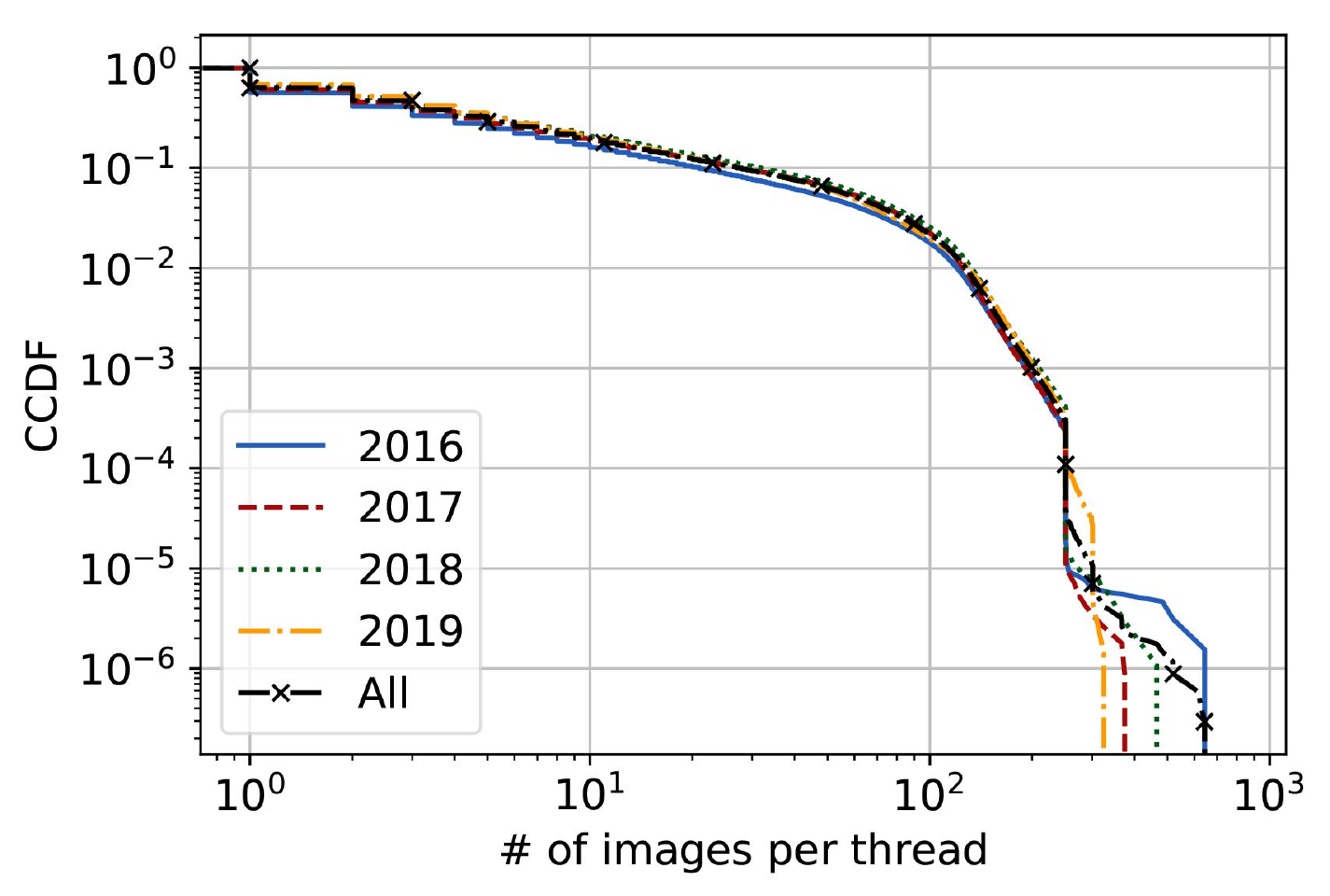}\label{fig:ccdf_pictures_per_thread_per_year}}
\caption{CDF and CCDF of the number of images per thread.}
\label{fig:cdf_ccdf_pictures_per_thread}
\end{figure}

\descr{Images.} Sharing images is very common on 4chan, in fact, OPs need to post an image when creating new threads.
Specifically, 4chan is mentioned by popular press and academic studies about the amount of original content (e.g., memes) it creates and disseminates across the Web~\cite{nyt2018meme,zannettou2018origins,vice2015meme}.
We aim to provide an overview of how many image metadata are included in our dataset.
To shed light on the use of images on \dspol over the years, we plot the CDF and CCDF of the number of images per thread in Figure~\ref{fig:cdf_ccdf_pictures_per_thread}.
We find that around $27\%$ of posts (36.9M) in our dataset include an image.
On average, 9.2, 10.8, 11.9, and 11 images appear, per thread, in 2016, 2017, 2018, and 2019, respectively.
Overall, 2017 was the year with the highest number of images shared: 12.1M.
Specifically, $17\%$ of the threads in 2016 have over 10 images, rising to $19\%$ in 2017, and eventually around $20\%$ in 2018 and 2019.
We test for statistically significant differences between the distributions using a two-sample KS test, and find them on each pair ($p<0.01$).

\section{Content Analysis}\label{sec:contentanalysis}
In this section, we provide an analysis of the content of the posts in our dataset.
More specifically, we detect the most popular topics discussed over the years, the named entities mentioned in each post, and how toxic a post is.

While the latter two are included in our data release, the first is not because topic extraction is done over {\em sets} of posts.
Nonetheless, we present it here to give an overview of what is discussed on \dspol, and thus is in the dataset.

\begin{table*}[t]
\centering
\resizebox{1.0\textwidth}{!}{
\begin{tabular}{r l}
\toprule
\textbf{Topic} & \textbf{Year 2016} \\
\midrule
1 & people (0.007), like (0.005), think (0.005), right (0.004), thing (0.004), know (0.004), 
polite (0.004), need (0.003), want (0.003), human (0.003) \\
2 & Trump (0.03), vote (0.021), elect (0.013), leaf (0.012), president (0.012), Hillary (0.011), fuck (0.01), shit (0.01), lose (0.009), happen (0.009) \\
3 & white (0.023), bump (0.013), nigger (0.013), country (0.009), praise (0.009), 
black (0.009), check (0.008), race (0.008), fuck (0.008), people (0.007) \\
4 & thread (0.022), Jew (0.014), fuck (0.014), faggot (0.014), good (0.011), kike (0.010), 
wrong (0.009), kill (0.009), shill (0.009), retard (0.009)\\
5 & fuck (0.009), girl (0.009), women (0.009), like (0.008), dick (0.007), cuck (0.007),
love (0.006), look (0.006), woman (0.006), lmao (0.006)\\
\toprule
\textbf{Topic} & \textbf{Year 2017} \\
\midrule
1 & post (0.021), shit (0.012), know (0.011), fuck (0.01), think (0.009), meme (0.009), 
retard (0.009), fake (0.008), mean (0.007), leaf (0.007) \\
2 & good (0.009), moor (0.008), lmao (0.006), base (0.006), go (0.006), kill (0.005), 
movie (0.004), fuck (0.004), like (0.004), roll (0.004) \\
3 & people (0.006), like (0.005), think (0.004), thing (0.004), work (0.003), want (0.003), 
know (0.003), right (0.003), social (0.003), human (0.003) \\
4 & nigger (0.012), fuck (0.007), money (0.006), like (0.006), people (0.006), year (0.005), 
work (0.005), want (0.005), live (0.005), shoot (0.005) \\
5 & thank (0.027), anon (0.021), kike (0.012), love (0.01), remind (0.008), fuck (0.008), 
maga (0.007), delete (0.007), sorry (0.007), time (0.007) \\
\toprule
\textbf{Topic} & \textbf{Year 2018} \\
\midrule
1 & bump (0.025), good (0.018), thank (0.017), anon (0.015), happen (0.01), Christmas (0.009), 
suck (0.007), dick (0.006), feel (0.006), hope (0.006) \\
2 & white (0.016), Jew (0.01), country (0.009), American (0.006), German (0.006), 
fuck (0.006), people (0.006), America (0.006), Europe (0.006), European (0.006) \\
3 & kike (0.024), right (0.014), fuck (0.012), mean (0.011), Israel (0.011), wall (0.01),
btfo (0.01), boomer (0.009), go (0.008), haha (0.007) \\
4 & money (0.007), work (0.007), year (0.006), people (0.006), live (0.005), like (0.004),
fuck (0.004), need (0.004), go (0.004), want (0.004) \\
5 & fuck (0.027), post (0.02), thread (0.019), faggot (0.013), shit (0.012), retard (0.01), 
know (0.01), shill (0.009), flag (0.009), meme (0.008) \\
\toprule
\textbf{Topic} & \textbf{Year 2019} \\
\midrule
1 & people (0.006), christian (0.006), believe (0.005), Jew (0.005), like (0.005), 
think (0.005), Jewish (0.004), know (0.004), read (0.004), white (0.004) \\
2 & white (0.015), country (0.009), Jew (0.009), America (0.007), American (0.007), 
china (0.006), people (0.006), Israel (0.006), fuck (0.006), Europe (0.005) \\
3 & fpbp (0.007), sage (0.007), drink (0.007), glow (0.006), nigga (0.006), like (0.005), 
fuck (0.005), tulsi (0.005), water (0.005), meat (0.005) \\
4 & base (0.089), bump (0.05), post (0.022), true (0.016), incel (0.015), cringe (0.014), 
redpill (0.014), know (0.012), seethe (0.011), btfo (0.01) \\
5 & kike (0.025), flag (0.024), nice (0.022), leaf (0.015), shill (0.015), meme (0.013), 
fuck (0.013), cope (0.011), memeflag (0.009), forget (0.008) \\
\toprule
\end{tabular}
}
\caption{Topics discussed on \dspol per year.}\label{tbl:yeartopics}
\end{table*}

\subsection{Topics} 
Looking at topics frequently mentioned on \dspol over the years provides a high-level reflection of the nature of discussions taking place on the board.
Importantly, researchers interested in studying discussions around specific topics included in this analysis can find our dataset useful.

We use Latent Dirichlet Allocation (LDA), which is used for basic topic modeling~\cite{blei2003latent}.
First, for each year, we collect the escaped HTML text provided for each post by the 4chan API.
Then, before tokenizing every post, we remove any stopwords, URLs, and HTML code.
Last, we create a term frequency-inverse document frequency (TF-IDF) array that is used to fit our LDA model.
TF-IDF statistically measures how important a word is to a collection of words; previous work shows it yields more accurate topics~\cite{mehrotra2013improving}.

In Table~\ref{tbl:yeartopics}, we list the top five topics discussed on \dspol for each year, along with the weights of each word for that topic.
We find that, during 2016, \dspol users were discussing political matters in a significant manner, and in particular the 2016 US Presidential Elections (topic 2).
We also find several topics with racist connotations, like \emph{kike} (derogatory term to denote Jews) and \emph{nigger}.
Other racist topics appear in other years as well, which highlights that controversial and racist words are used frequently on \dspol.

Overall, our topic analysis shows that discussions in \dspol feature political matters, hate, misogyny, and racism over the course of our dataset.

\subsection{Toxicity}
Next, we set to score the content of the posts according to how toxic, inflammatory, profane, insulting, obscene, or spammy the text is.
To this end, we use Google's Perspective API~\cite{jigsaw2018perspective}, which offers several models for scoring text trained over crowdsourced annotations.
We choose Google’s Perspective API as other available methods mostly use short texts (tweets) for their training samples~\cite{davidson2017automated}. 
Perspective API should perform better for our dataset as it was trained using comments with no restriction in character length~\cite{wired2017api}, similar to the comments of our dataset.

We focus on the following 7 models:
\begin{compactitem}[--]
  \item {\sc toxicity} and {\sc severe\_toxicity}: quantify how rude or disrespectful a comment is; note that the latter is less sensitive to messages that include positive uses of curse words compared to the former.
  \item {\sc inflammatory}: how likely it is for a message to ``inflame'' discussion.
  \item {\sc profanity}: how likely a message is to contain swear or curse words.
  \item {\sc insult}: how likely a message is to contain insulting or negative content towards an individual or group of individuals.
  \item {\sc obscene}: how likely a message is to contain obscene language.
  \item {\sc spam}: how likely a message is to be spam.
\end{compactitem}
We score each post in our dataset using the API and include the results in the final dataset.
We only obtain results for posts that include text, since scores are computed only over text.
That is, we do not score $2.3\%$ (3.1M) of the posts in our dataset that have no text. 

\begin{figure*}[t!]
\center
\subfigure[]{\includegraphics[width=0.27\textwidth]{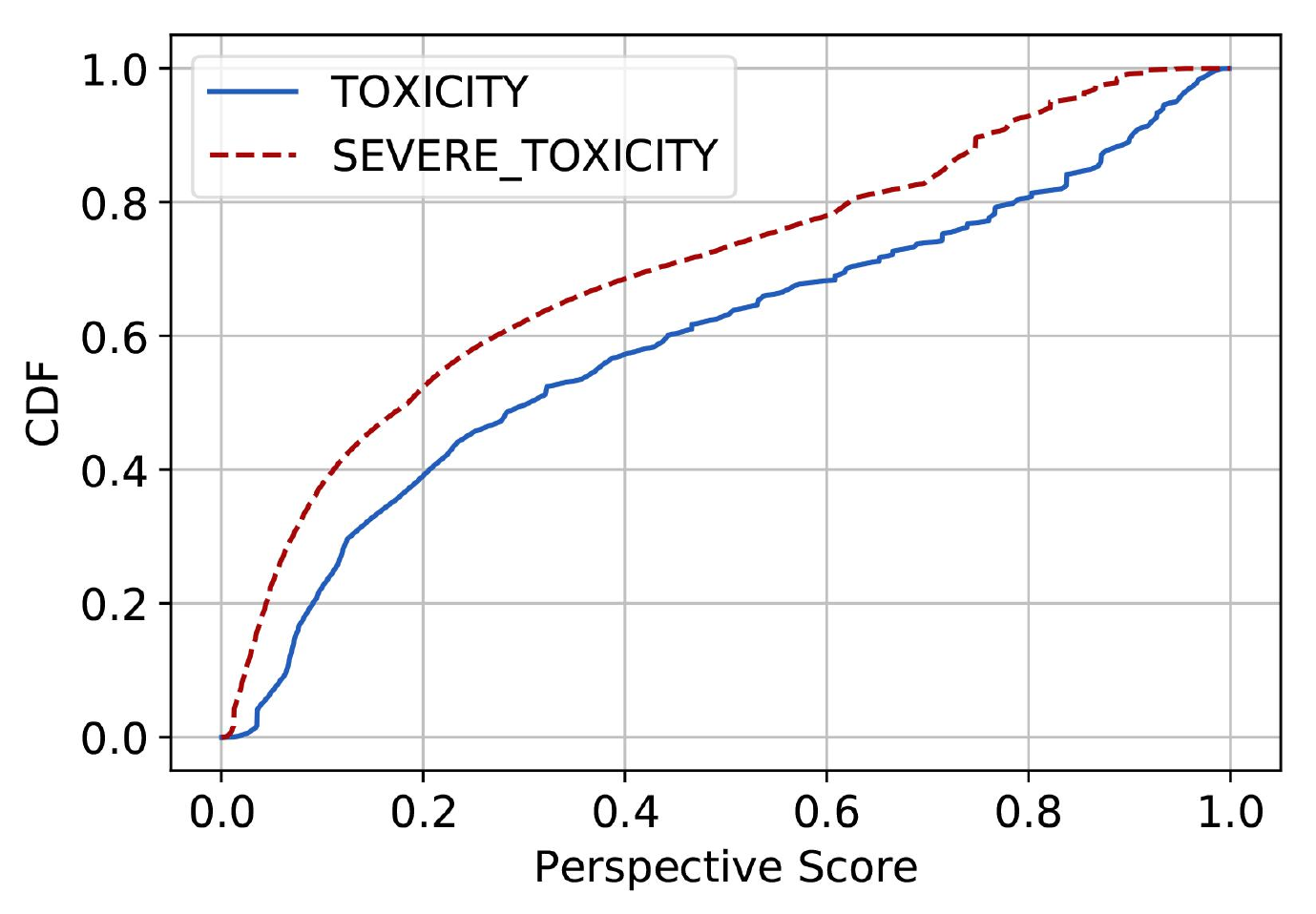}\label{fig:cdf_perspective_toxicity}}
\subfigure[]{\includegraphics[width=0.27\textwidth]{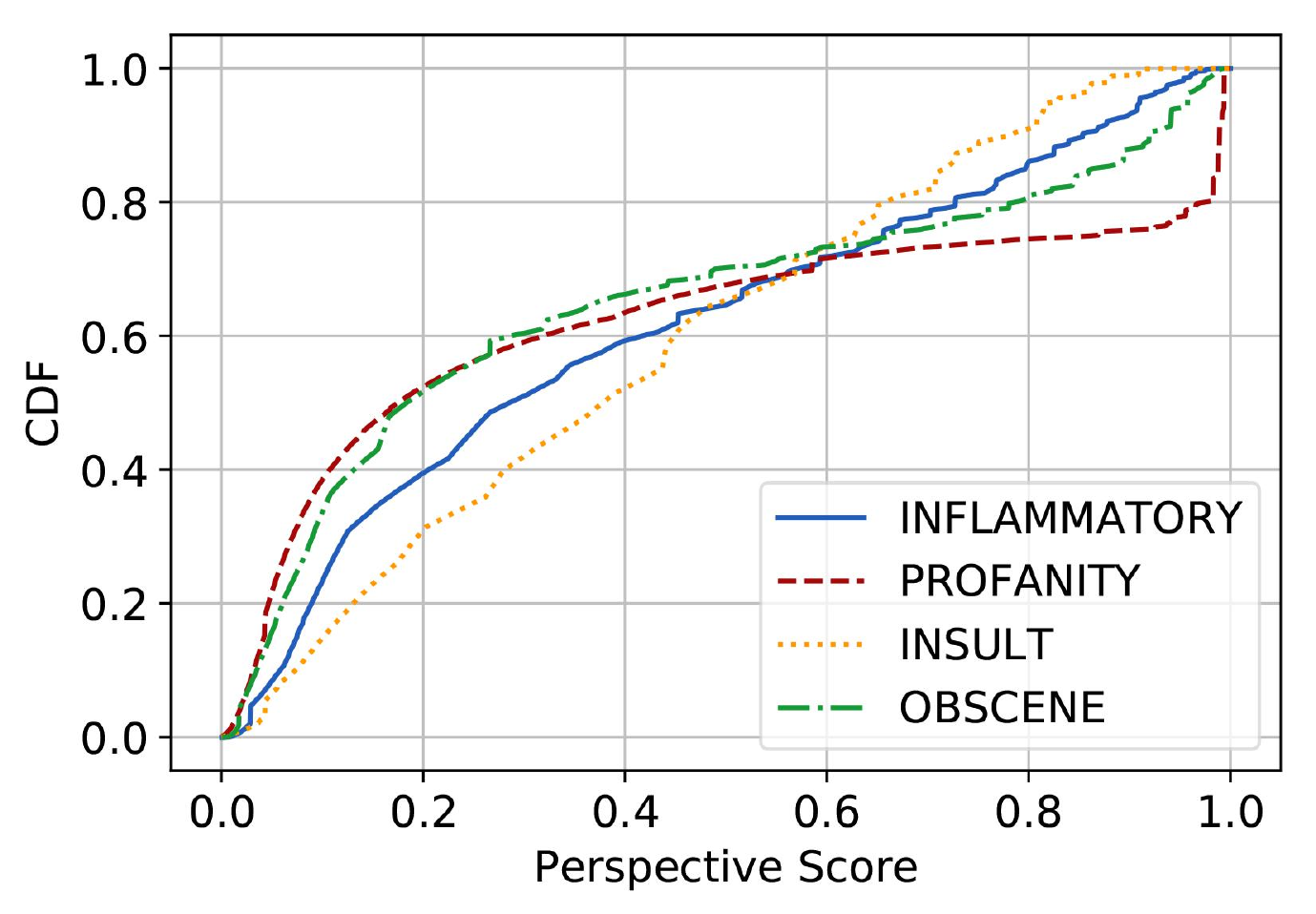}\label{fig:cdf_perspective_insult}}
\subfigure[]{\includegraphics[width=0.27\textwidth]{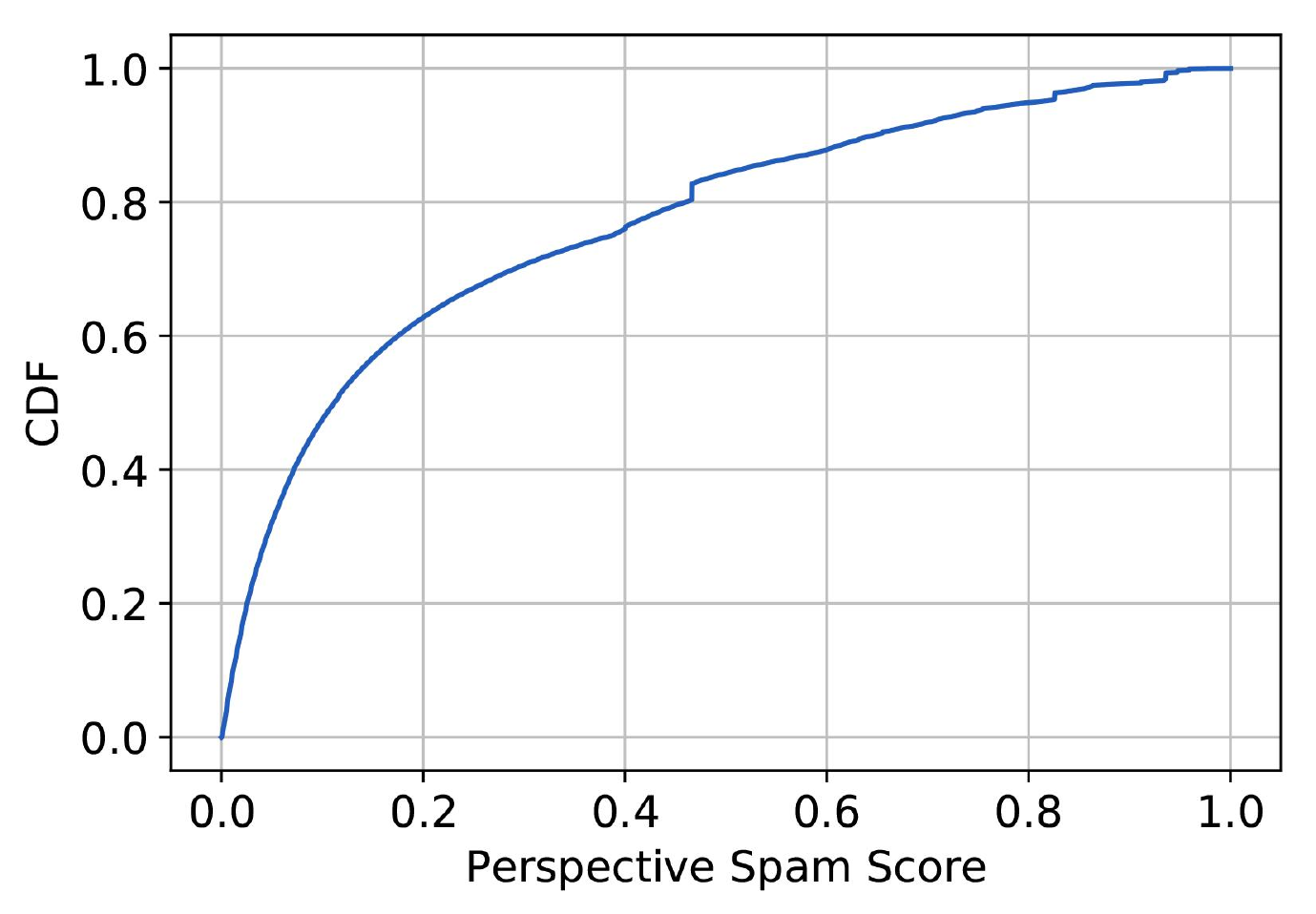}\label{fig:cdf_perspective_spam}}
  \caption{CDF of the Perspective Scores related to how toxic, inflammatory, obscene, profane, insulting, or spammy is a post. }
\label{fig:cdf_perspective_scores}
\end{figure*}

In Figure~\ref{fig:cdf_perspective_scores}, we plot the CDF of the scores for each of the models.
We observe that \dspol exhibits a high degree of toxic content: $37\%$ and $27\%$ of the posts have, respectively, {\sc toxicity} and {\sc severe\_toxicity} scores greater than 0.5 (see Figure~\ref{fig:cdf_perspective_toxicity}).
These results are in line with previous research findings~\cite{hine2017kek}.  
For the other models, we observe similar trends: $36\%$ of the posts have an {\sc inflammatory} score greater than 0.5 (Figure~\ref{fig:cdf_perspective_insult}), $33\%$ for {\sc profanity} (Figure~\ref{fig:cdf_perspective_insult}), $35\%$ for {\sc insult} (Figure~\ref{fig:cdf_perspective_insult}), $30\%$ for {\sc obscene} (Figure~\ref{fig:cdf_perspective_insult}), but only $16\%$ for {\sc spam} (Figure~\ref{fig:cdf_perspective_spam}).
We also test for statistically significant differences between the distributions in Figure~\ref{fig:cdf_perspective_scores}, using two-sample KS test, and find them on each pair ($p<0.01$). 

Overall, we are confident that this additional set of labels can be extremely useful for researchers studying hate speech, bullying, and aggression on the Web.

\subsection{Named Entity Recognition}

\begin{table}[t]

\centering
\resizebox{0.9\columnwidth}{!}{
\begin{tabular}{ l r r | l r r}
\toprule
\textbf{Named Entity} & \textbf{\#Posts} & \textbf{($\%$)} &\textbf{Entity Label} & \textbf{\#Posts} & \textbf{($\%$)}\\
\midrule
Trump & 2,461,452 & 1.83 & DATE & 92,945,374 & 69.06 \\
one & 1,811,983 & 1.35 & CARDINAL & 20,069,995 & 14.92 \\
first & 1,584,686 & 1.18 & PERSON & 17,532,857 & 13.03 \\
US & 1,066,408 & 0.79 & ORG & 17,145,386 & 12.74 \\
Jews & 963,398 & 0.72 & NORP & 16,820,469 & 12.50 \\
America & 831,007 & 0.62 & GPE & 14,813,739 & 11.01 \\
Europe & 719,873 & 0.54 & TIME & 4,498,824 & 3.34 \\
two & 703,767 & 0.52 & ORDINAL & 2,923,765 & 2.17 \\
American & 676,332 & 0.50 & LOC & 2,676,504 & 1.99 \\
Israel & 589,718 & 0.44 & PERCENT & 2,189,227 & 1.68 \\
\toprule
\end{tabular}
}
\caption{Top 10 named entity and entity label that appear in \dspol posts.}
\label{tbl:label_entity_occurrunces}
\end{table}

Finally, we extract the ``named entities'' mentioned in \dspol posts, as we hope this will allow the research community to study discussions around specific entities, e.g.,  individuals, countries, etc.
To obtain the named entities, we use the \emph{en\_core\_web\_lg} model publicly available via the SpaCy library~\cite{spacy.io}.
We choose this specific model over other alternatives since it was trained with the largest available dataset.
In addition, previous work~\cite{jiang2016evaluating} ranked it among the top two most accurate methods for named entity recognition.
It uses millions of Web entries consisting of news articles, blogs, and comments to detect and extract a variety of entities from text.
Entities range from specific popular individuals to nationalities, countries, and even events.\footnote{See \url{https://spacy.io/api/annotation\#named-entities} for the full list of labels.}

We run the entity detection model against all the posts in our dataset and include the extracted entities in the final dataset.
Note that the model did not return any entities for 18M posts ($13\%$); this is expected since a lot of posts do not reference any entities and due to the fact that a considerable number of posts do not have any text.

In Table~\ref{tbl:label_entity_occurrunces}, we list the ten most popular named entities in our dataset.
Note that a post can mention a popular entity more than once.
We report the number of posts in our dataset that mention an entity {\em at least} once.
We find that Donald Trump is the most popular named entity on \dspol with over 2.46M posts ($1.83\%$) mentioning him.
Other popular named entities include ``US'' ($0.79\%$), ``Jews'' ($0.72\%$), ``America'' ($0.62\%$), ``Europe'' ($0.54\%$), ``American'' ($0.50\%$), and ``Israel'' ($0.44\%$).
We also report the top ten entity {\em labels} in our dataset.
The entity labels specify the category of the entity mentioned in each post (e.g., ``PERSON'' for Donald Trump).
The most popular label is date ($69.06\%$), followed by cardinal numbers ($14.92\%$), and real or imaginary people ($13.03\%$).
Other popular labels include organizations ($12.74\%$), nationalities, religious, or political groups ($12.50\%$), and times smaller than a day ($3.34\%$).
Reviewing the most popular named entities and labels of our dataset suggests that discussions on \dspol are related to discussions about world happenings and events.

Overall, we hope that augmenting our dataset with the named entities will be valuable to researchers working on Computational Social Sciences who wish to study discussions around specific individuals, nationalities, etc.

\section{Related Work}\label{sec:relatedwork}
In this section, we review relevant related work.
Over the past couple of years, a number of research papers have used data collected from 4chan; some also mention that data is available upon request.
Overall, our 4chan dataset is, to the best of our knowledge, 1) the only one to be freely and publicly available online, and 2) the largest and most comprehensive one, including 3.5 years worth of data.

\descr{Studies focusing on 4chan.} Bernstein et al.~\cite{bernstein20114chan} crawl 5.5M posts from 500K threads posted on the ``Random'' (/b/) board between July 19 and August 2, 2010, and present a content analysis showing how posts are dominated by images and posting of external URLs.
Their dataset is not openly accessible.
Hine et al.~\cite{hine2017kek} collect 11M posts from June 30 to September 12, 2016 from 3 different boards, namely, ``Politically Incorrect'' (\dspol), ``Sports'' (/sp/), and ``International'' (/int/), presenting a general characterization of the former while mostly using the latter two for comparison.
Overall, they study the effect of ephemerality and bump limits, and show that \dspol is characterized by a high degree of hate speech.
Moreover, they find that the board serves as an aggregation point for {\em coordinated} harassment campaigns on other platforms such as YouTube.
Given the timeline of the data (Summer 2016), a lot of the content is related to the 2016 US Presidential Election, with 4chan users exhibiting unconventional support, often in terms of memes and novel image content, to Donald Trump's 2016 presidential campaign.
The dataset of this study is only available upon request and, more importantly, only includes 2.5 months rather than 3.5 years worth of data.

Tuters and Hagen~\cite{tuters2019they} analyze 1M posts from 4chan's \dspol that contained words enclosed in triple parenthesis, i.e., ((())).
They find that such posts often feature anti-Semitic nature and that \dspol posters tend to create and use political and racist memes.
This dataset is not openly accessible.

Finally, Pettis~\cite{pettisambiguity} collect 2.7K and 1.1K threads from \dspol and the ``Technology'' board (/g/), respectively and focus on qualitatively studying whether anonymity lets individuals be more open to reveal their emotions and beliefs online.
Again, this dataset is not available online.

\descr{Multi-platform studies.} Zannettou et al.~\cite{zannettou2017web} study how mainstream and fringe Web communities (4chan, Reddit, and Twitter) share mainstream and alternative news sources to influence each other.
Between June 30, 2016 and February 28, 2017 they collected: a) 487K tweets; b) 42M posts, 390M comments, and 300K subreddits; and c) 97K posts made on \dspol, /sp/, /int/, and the ``Science'' board (/sci/).
They find that, before a story is made popular, it was often posted on 4chan for the first time, and use a statistical method called Hawkes Process to quantify the influence of 4chan with respect to news dissemination.
This dataset is available upon request.
Snyder et al.~\cite{snyder2017fifteen} collect more than 1.45M posts from paste-bin.com, 282K posts from \dspol and /b/, and 4K posts from 8ch's \dspol and /baphomet/ to detect doxing. 
This dataset is not publicly available.
Then, Zannettou et al.~\cite{zannettou2018origins} present a large-scale measurement study of the meme ecosystem, using 160M images obtained from \dspol, Reddit, Twitter, and Gab.
They collect 74M unique images from Twitter, 30M from Reddit, 193K from Gab, and 3.6M from \dspol.
The study shows that Reddit and Twitter tend to post memes for ``fun,'' while Gab and \dspol users post racist and political memes targeting specific audiences.
Importantly, they find that \dspol is the leading creator of racist and political memes, and the subreddit "The\_Donald" is very successful in disseminating memes to both fringe and mainstream Web communities.
The authors created an openly accessible dataset, however, it only consists of the URLs and the hashes of the images collected.
Finally, Mittos et al.~\cite{mittos2020and} gather 1.9M threads from \dspol, along with the pictures posted, and 2B comments from 473K subreddits. 
They extract posts that might be related to genetic testing, showing the context in which genetic testing is discussed and finding that it often yields high user engagement. 
In addition, the discussion of this topic often includes hateful, racist, and misogynistic comments. 
Specifically, \dspol conversations about genetic testing involves several alt-right personalities, antisemitism, and hateful memes.
The authors did not make their dataset openly accessible.

\descr{Dataset Papers.} Here we list other dataset papers that are also somewhat related to the motivations behind our work, in that they release data associated with social network content as well as potentially nefarious activities.
Brena et al.~\cite{brena2019news} present a data collection pipeline and a dataset with news articles along with their associated sharing activity on Twitter, which is relevant in studying the involvement of Twitter users in news dissemination.
The pipeline can also be used to classify the political party supported by Twitter users, based on the news outlets they share along with the hashtags they post on their tweets.
Fair and Wesslen~\cite{fair2019shouting} present a dataset of 37M posts, 24.5M comments, and 819K user profiles collected from the social network Gab, which, like 4chan, is often associated to alt-right and hateful content.
Their dataset includes user account data, along with friends and follower information, and edited posts and comments in case a user made an edit.

Garimella and Tyson~\cite{garimella2018whatapp} present a methodology for collecting large-scale data from WhatsApp public groups and release an anonymized version of the collected data.
They scrape data from 200 public groups and obtain 454K messages from 45K users.
They analyze the topics discussed, as well as the frequency and topics of the messages to characterize the communication patterns in WhatsApp groups.
Finally, Founta et al.~\cite{founta2018crowdsourcing} use crowdsourcing to label a dataset of 80K tweets as normal, spam, abusive, or hateful.
More specifically, they release the tweet IDs (not the actual tweet) along with the majority label received from the crowdworkers.

\section{Conclusion}\label{sec:conclusion}
This paper presented our 4chan dataset; to the best of our knowledge, the largest publicly available dataset of its kind.
The dataset includes over 3.3M threads and 134.5M posts from 4chan's Politically Incorrect board collected between June 2016 and November 2019.
We also augmented the dataset with a set of labels measuring the toxicity of each post, as well as the named entities mentioned in each post.

Overall, we are confident that our work will further motivate and assist researchers in studying and understanding 4chan as well as its role on the greater Web.
Access to the dataset could also help answer numerous questions about \dspol, 
e.g., what is the nature of discussion on the board following sharing of news articles?
what is the role played by 4chan in alternative and fake news dissemination? 
what is 4chan's role in coordinated aggression campaigns, doxing, trolling, etc.?
Moreover, using this dataset in conjunction with data from other social networks could also help researchers understand the similarities and differences of users of different communities. 
Also, our dataset is an invaluable resource for training algorithms in natural language processing, modeling of slang words, or detecting hate speech, fake news dissemination, conspiracy theories, etc.
Finally, we hope that the data can be used in qualitative work to present in-depth case studies of specific narratives, events, or social theories.

\descr{Acknowledgments.}
This work was funded by the EU Horizon 2020 Research and Innovation program under the Marie Skłodowska Curie ENCASE project (GA No.~691025), 
the US National Science Foundation (Grant No.~CNS-1942610), 
and the UK EPSRC grant EP/S022503/1 that supports the 
Centre for Doctoral Training in Cybersecurity.

\small
\bibliographystyle{abbrv}
\bibliography{references}

\end{document}